\documentclass[pre,preprint,showpacs]{revtex4}% Physical Review B

\usepackage{graphicx}
\usepackage{amsmath,amssymb}
\begin{document}
\title{Wavelength selection of rippling patterns in myxobacteria }
\author{L.L. Bonilla$^1$, A. Glavan$^1$, and A. Marquina$^2$}
\affiliation {$^1$G. Mill\'an Institute, Fluid Dynamics, Nanoscience and Industrial
Mathematics, Universidad Carlos III de Madrid, Avda.\ Universidad 30; E-28911 Legan\'es, Spain\\
$^2$Departmento de Matem\'atica Aplicada, Universidad de Valencia, Avda. Dr. Moliner 50; E-46100 Burjassot-Valencia, Spain}
\date{\today}
\begin{abstract}
Rippling patterns of myxobacteria appear in starving colonies before they aggregate to form fruiting bodies. These periodic traveling cell density waves arise from the coordination of individual cell reversals, resulting from an internal clock regulating them, and from contact signaling during bacterial collisions. Here we revisit a mathematical model of rippling in myxobacteria due to Igoshin et al.\ [Proc. Natl. Acad. Sci. USA {\bf 98}, 14913 (2001) and Phys. Rev. E {\bf 70}, 041911 (2004)]. Bacteria in this model are phase oscillators with an extra internal phase through which they are coupled to a mean-field of oppositely moving bacteria. Previously, patterns for this model were obtained only by numerical methods and it was not possible to find their wavenumber analytically. We derive an evolution equation for the reversal point density that selects the pattern wavenumber in the weak signaling limit, show the validity of the selection rule by solving numerically the model equations and describe other stable patterns in the strong signaling limit. The nonlocal mean-field coupling tends to decohere and confine patterns. Under appropriate circumstances, it can annihilate the patterns leaving a constant density state via a nonequilibrium phase transition reminiscent of destruction of synchronization in the Kuramoto model. 
\end{abstract}
\maketitle 

\section{Introduction}
\label{sec:1}

Self-organized patterns are ubiquitous in bacterial colonies \cite{sek03}. Cooperative behavior is often produced by motion following chemical (chemotaxis) or adhesive (haptotaxis) gradients and modeled using reaction-diffusion equations \cite{murray}. Instead of using diffusion of chemicals to communicate cells and produce patterns, direct cell contacts yield density waves in myxobacteria such as \textit{Myxoccocus xanthus} \cite{igo01,igo04,igo04b,sozi05,wu09}. Myxobacteria are rod-shaped gram-negative bacteria, components of soil, that move by gliding in the direction of their long axis on solid surfaces, either as individuals or in swarms \cite{hen72}. This type of motility of \textit{M. xanthus} cells is controlled by two different motors concentrated at the cell poles: the A-motility system (adventurous) specific for individual cell motion and the S-motility system (social), for group cell motion. The A-engine is a `pusher' and works by secretion and hydration of slime, a polyelectrolyte gel, whereas the S-engine is a `puller' and operates through extension and retraction of type IV pili \cite{wol02}. 

When nutrients are abundant, myxobacteria aggregate into multicellular swarms and spread outwards from the border of the colony. However they respond to starvation conditions by reversing their directions, moving towards the colony center and forming a multicellular fruiting body with nonmotile spores. Spores can survive for long periods of time and, under adequate conditions, germinate giving rise to motile vegetative cells \cite{kim90a}. Fruiting body formation is a multi-step process of alignment, rippling, streaming, and aggregation \cite{kai10}. Overall models of the myxobacteria life cycle depending on food availability can be found in \cite{sozi05,alb04,hen10}. Before aggregation begins, bacterial collective behavior and intercellular communication create fascinating collective concentric and spiral traveling waves called {\em ripples}. During the rippling phase, bacteria move in concert in such a way that colliding waves appear to pass through one another. The resulting periodic patterns consist of equally spaced ridges of high cell density that appear to advance through the bacterial population as rhythmically traveling waves moving in opposite directions. In these counter-propagating waves, individual cells are parallel to the direction of the ripples, move back and forth and exchange developmental signals (C-signals) only when they collide \cite{kim90b,jel03,sag94}. Most experimentally observed rippling patterns can be characterized as counter-propagating traveling waves \cite{wel01,igo01}. Unlike waves generated by reaction-diffusion instabilities, which annihilate on collision, myxobacteria waves appear to pass through one another unaffected and are therefore  analogous to solitons \cite{igo04}.

There are discrete \cite{bor02,alb04,and05,sli06,wu07} and continuum \cite{igo01,igo04,lut02,bor04,gal06} theoretical models of cell behavior and cell-to-cell interaction based on C-signaling. Intercellular communication is by direct cell contact, without any diffusible morphogen signaling. Cells migrating in opposite directions can come into contact with each other (collide) and exchange C-signals. Cell state changes periodically so that isolated cells reverse periodically their motion. These changes can be modeled by the concept of an internal clock \cite{igo01,igo04,wu09}. The internal clock or pacemaker regulates swarming that is driven by growth \cite{kai11}. In some models rippling cannot occur unless there is a refractory period after cell reversal during which a cell does not respond to C-signals from other cells and does not reverse \cite{bor02,igo01,igo04}. Rippling may occur even without a refractory phase in generalized discrete models if the interaction strength and the degree of cooperativity are large enough \cite{bor04}. 

In this paper, we consider Igoshin et al's continuum one-dimensional (1D) model for rippling \cite{igo01,igo04}, solve it numerically and interpret the results by a study of the small nonlinearity limit. An extension of this model to 2D can be found in \cite{igo04b} and related agent-based models in \cite{zha12}. Previous work has shown that numerical solutions of the continuum model exhibit rippling \cite{igo04}. Surprisingly, the analysis of the small nonlinearity limit does not provide a selection rule for the observed wavelength of ripples \cite{igo04}. In contrast with these results, we show that the small nonlinearity limit does provide a description of the rippling instability and it supplies the critical wavenumber of the periodic patterns in terms of the refractory period. In fact, careful consideration of flux continuity at the different stages of one cycle of the internal clock of myxobacteria shows that the limiting equation for the cell density is of Fokker-Planck type but it contains an additional source term. The extra source term is the cell density times a nonlocal growth rate that vanishes as the refractory period tends to zero. The source term is key new element in our analysis, as it produces dissipation even in the absence of diffusion. Thus the role of the source term is similar to that of collision terms in the Boltzmann transport equation. The balance between advection in space and time and dissipation due to the ``collision term'' selects the wavelength of the rippling patterns that issue forth from the uniform stationary state. These patterns are periodic in time and space. Numerical solutions of the full model equations show that the pattern frequency decreases as the strength of the nonlinearity increases. 

It is interesting to contrast the behavior of the present myxobacteria model with the well-known synchronization of globally coupled phase oscillators described by the Kuramoto model \cite{kuramoto,rmp05}. The Kuramoto oscillators move on a circle with their own random natural frequency and their mutual interaction through a mean field tends to synchronize them. Partial or complete synchronization is achieved for sufficiently strong coupling through a nonequilibrium phase transition. Igoshin et al's model describes phase oscillators with an internal clock whose angular speed changes according to their interaction with oppositely moving oscillators. Since the Igoshin oscillators move with a constant positive or negative velocity, patterns arising from appropriate initial conditions persist in the absence of interaction. In this case, mean-field interaction among oscillators may produce loss of rhythmicity resulting in the destruction of patterns. This change appears as a nonequilibrium phase transition at critical values of parameters that has been explicitly shown in the weak coupling limit.

The rest of the paper is as follows. We describe Igoshin et al's continuum model \cite{igo04} in Section \ref{sec:2} and write the corresponding equations in nondimensional form. The weak signaling limit is examined in Section \ref{sec:3}, where the limiting Fokker-Planck type equation with source term is derived. The linear stability of its uniform stationary solution is analyzed in Section \ref{sec:4}. In the absence of diffusion and for disturbances with frequency one (same as that of the signaling solution without nonlinear terms), we find that rippling patterns appear for disturbances with wavenumber less than one, whatever the refractory period. Increasing the wavenumber $k$ may result in the cancellation of patterns, as the uniform stationary solution becomes linearly stable. Since diffusion has a stabilizing role for the uniform solution (producing a negative term proportional to $k^2$ in the real part of eigenvalues), unavoidable numerical noise should tend to annihilate the periodic patterns and be more effective as $k$ increases. Section \ref{sec:5} contains the results of numerical solutions of the full model equations. The findings of the linear stability analysis and the wavenumber selection criteria are confirmed. In addition, we observe a rich variety of stable periodic patterns comprising standing and traveling waves as numerical solutions of the model equations for moderate and strong signaling between cells. Section \ref{sec:6} contrasts our findings with the behavior of the Kuramoto model for the synchronization of phase oscillators. The last section contains our conclusions.

\setcounter{equation}{0}
\section{Model}
\label{sec:2}

When starving, myxobacteria aggregate and form a fruiting body and eventually a spore. Before aggregation begins, there appear periodic patterns of equally spaced high density bands that move as traveling waves (ripples). The one-dimensional (1D) model for ripples of Igoshin et al \cite{igo01,igo04} is based on the following experimental observations:

- \textit{Internal biochemical clock}: cells are aligned parallel and glide along their long axis in one direction during part of their internal period and glide in the opposite direction during the other part of their period.

- \textit{Contact signaling}: a cell collides with an oppositely moving cell and interchanges a signal (a C-protein). As a result, both cells reverse their motion. The collision frequency depends on the local cell density.

- \textit{Refractory period}: after one collision there is a refractory period during which the cell does not reverse its motion even if it collides again.

Let $n(x,\phi,t)$ be the number density of bacteria per unit length $x$ and per unit internal phase $\phi$. Cells with $0<\phi<\pi$ move to the right with velocity $v$ and cells with $-\pi<\phi<0$ move to the left (velocity $-v$). Cell division and death are negligible during the rippling phase. The governing equation is as follows:
\begin{eqnarray}
&& \partial_t n+ v(\phi)\partial_x n-D_x\partial_x^2n+ \partial_\phi J(x,\phi,t) = 0, \label{eq1}\\
&& v(\phi)= v\,\mbox{sign}\phi. \label{eq2}
\end{eqnarray}
Here $D_x\geq 0$ is a space diffusivity, the density $n(x,\phi,t)$ is a $2\pi$-periodic function of $\phi$ and it satisfies periodic boundary conditions at $x=\pm L$ or it decays to zero if $L=\infty$. To model the angular flux $J(x,\phi,t)$, we make precise the above experimental observations \cite{igo04}. The cell velocity is given by (\ref{eq2}) and reversals occur at $\phi=0,\pm\pi$. The internal clock of a bacterium advances with constant angular velocity $\omega$ but when the cell collides with another one moving in the opposite direction, both cells exchange a signal. The collision frequency, and therefore the signaling intensity, is proportional to the local cell density. The cells may respond positively to this signal by accelerating their angular speed from $\omega$ to $\omega+\epsilon \omega \Omega(N_\pm)$ (see below) depending on their internal phase. The coefficient $\epsilon$ measures the relative change in angular velocity from that during the refractory period, $\omega$, to the bacterium velocity during the signaling period, thereby characterizing signaling strength. After each reversal (at $\phi=0, \pm\pi$), the cell enters a refractory period during which does not respond to collision signaling and does not reverse. If $\alpha>0$ is the angular duration of the refractory period, the angular flux is $J=\omega\, n(x,\phi,t)$ for $0<\phi<\alpha$ and for $-\pi<\phi<-\pi+\alpha$ ($0\leq\alpha\leq\pi$). Overall the angular flux is \cite{igo04} 
\begin{eqnarray}
&&J= \omega\, n\, [1+ \epsilon\Omega(N_-(x,t))\chi_{[\alpha,\pi]}(\phi) + \epsilon \Omega(N_+(x,t))\chi_{[-\pi+\alpha,0]}(\phi)]-D_\phi \partial_\phi n, \label{eq3}\\
&&\chi_{[A,B]}(\phi)= \left\{ \begin{array}{cc} 
1, & A<\phi<B,\\ 
0, & \mbox{otherwise},
 \end{array}\right.     \label{eq4}
\end{eqnarray}
where 
\begin{equation}
N_+(x,t)= \int_0^\pi n(x,\phi,t)\, d\phi,\quad N_-(x,t)= \int_{-\pi}^0 n(x,\phi,t)\, d\phi,      \label{eq5}
\end{equation}
$D_\phi$ is a positive number and 
\begin{equation}
\Omega(N)= \frac{N^r}{N^r+N^r_{\rm cr}},      \label{eq6}
\end{equation}
with $r>0$. We shall use $r=4$ \cite{igo04}. In Ref.\ \onlinecite{igo04}, it is explained that both $D_x$ and $D_\phi$ are small. The total density of bacteria at point $x$ and time $t$ is $N_+(x,t)+N_-(x,t)$ and the density of time-reversal points ({\em reversal point density}) is 
\begin{equation}
n_{\rm RPD}(x,t)=n(x,\phi=0+,t)+n(x,\phi=-\pi+,t). \label{RPD}
\end{equation}
Here $n(x,\phi=0+,t)$ and $n(x,\phi=-\pi+,t)$ are, respectively, the density of left-to-right and of right-to-left reversal points in spacetime. Thus their sum, as in \eqref{RPD}, is the density of all reversal points in spacetime.

The total number of myxobacterium cells should be independent of time. This means that $dN/dt=0$ in (\ref{eq1}), where
\begin{eqnarray}
N=\int_{-L}^L \int_{-\pi}^\pi n(x,\phi,t)\, dx d\phi.  \label{eq7}
\end{eqnarray}
In (\ref{eq1}), $n$ and $\partial_\phi n$ are continuous and $2\pi$-periodic in $\phi$, so that $N$ is independent of time if and only if
\begin{eqnarray}
\int_{-L}^L \left([J]_{\phi=-\pi}^{-\pi+\alpha}+[J]_{\phi=-\pi+\alpha}^0+[J]_{\phi=0}^\alpha + [J]_{\phi=\alpha}^\pi\right) dx=0.  \label{eq8}
\end{eqnarray}
Here we have used the boundary conditions at $x=\pm L$ (including the case $L=\infty$) and the notations $[f(x,\phi)]_{\phi=-\pi}^{-\pi+\alpha} = f(x,-\pi+\alpha)-f(x,-\pi)$ and so on. For the angular flux (\ref{eq3}), (\ref{eq8}) becomes
\begin{eqnarray}
\int_{-L}^L \{\Omega(N_+)\, [n]_{\phi=-\pi+\alpha}^0+\Omega(N_-)\,  [n]_{\phi=\alpha}^\pi\}\, dx= 0.  \label{eq9}
\end{eqnarray}

If there is no refractory period so that $\alpha=0$, $2\pi$-periodicity of $n$ and $\partial_\phi n$ yields
\begin{eqnarray}
\int_{-L}^L [\Omega(N_+)-\Omega(N_-)]\, [n]_{\phi=0}^\pi\, dx =0,  \label{eq10}
\end{eqnarray}
instead of (\ref{eq9}).

\begin{table}[ht]
\begin{center}\begin{tabular}{cccc}
 \hline
 $t$ & $x$ & $\phi$ &$n, N_\pm$   \\
$\frac{1}{\omega}$ & $\frac{v}{\omega}$ & $1$ &$N_{\rm cr}$ \\
 \hline
\end{tabular}
\end{center}
\caption{Nondimensional units.}
\label{t1}
\end{table}

It is convenient to render the equations of the model dimensionless. To this purpose, we shall use the units $[t]=1/\omega$, $[x]=v/\omega$, etc listed in Table \ref{t1}. Let us define dimensionless variables as $\hat{t}=t/[t]$, $\hat{x}=x/[x]$, and so on. Inserting these definitions in (\ref{eq1}), (\ref{eq3}) and (\ref{eq7}) and dropping hats in the results, we obtain the following equations:
\begin{eqnarray}
&&\partial_t n\!+\! \mbox{sign}(\phi)\partial_x n\! + \!\partial_\phi n\!= \!\epsilon\!\left\{\frac{\mathcal{D}_x}{2\pi}\partial_x^2n\!+\! \frac{\mathcal{D}_\phi}{2\pi}\partial_\phi^2 n\! -\!\partial_\phi([\Omega(N_-)\chi_{[\alpha,\pi]}\!+\!\Omega(N_+)\chi_{[-\pi+\alpha,0]}]n)\! \right\}\!\!, \label{eq11}\\
&&\int_{-\mathcal{L}}^\mathcal{L}\int_{-\pi}^\pi n(x,\phi,t)\, dx d\phi = \int_{-\mathcal{L}}^\mathcal{L} [N_+(x,t)+N_-(x,t)]\, dx = \hat{N}, \label{eq12}
\end{eqnarray}
where (the dimensionless parameters are assumed to be of order unity):
\begin{eqnarray}
 \mathcal{D}_x=\frac{2\pi D_x\omega}{\epsilon v^2}, \quad  \mathcal{D}_{\phi}=\frac{2\pi D_\phi}{\epsilon\omega}, \quad \mathcal{L} =\frac{\omega L}{v},\quad\hat{N}\equiv \frac{N\omega}{N_{\rm cr}v}, \quad \Omega(y)=\frac{y^r}{y^r+1}. \label{eq13}
\end{eqnarray}

For $\mathcal{D}_x=\mathcal{D}_\phi=0$, (\ref{eq11}) resembles a hyperbolic equation (or a system of two hyperbolic equations for oppositely moving bacteria). %with nonlocal advection terms that describe interaction between colliding bacteria by its effects on the internal clock thereof. 
However, strictly speaking this system is only hyperbolic in the spatial dimension. The phase fluxes (resulting from the nonlinear interaction between oppositely moving bacteria) are {\em nonlocal} in the clock angle: they are defined as integrals over the whole angular domain implying that the resulting 2D system (in $x$ and $\phi$) is integro-differential. The role of the nonlocal advection as generator of dissipation will be shown by the analysis of the weakly nonlinear limit and by solving \eqref{eq11} using a high order accurate weighted essentially non-oscillatory (WENO) numerical method \cite{jia96}.

\section{Weak signaling limit}
\label{sec:3}
For $\epsilon=0$, (\ref{eq11}) becomes
\begin{eqnarray}
\partial_t n+ \mbox{sign}(\phi)\partial_x n + \partial_\phi n = 0, \label{eq14}
\end{eqnarray}
whose solution is \cite{igo04}
\begin{eqnarray}
n= \left\{\begin{array}{cc}
f(x-\phi,t-\phi), & 0<\phi<\pi,\\
f(x+\phi,t-\phi), & -\pi<\phi< 0.\end{array}\right. \label{eq15}
\end{eqnarray}
Here $f(x,t)$ is an arbitrary function, $2\pi$-periodic in its second argument. $f(x,t)$ and $f(x-\pi,t+\pi)$ represent the densities of left-to-right and of right-to-left reversals, respectively. Then, according to \eqref{RPD}, $n_{\rm RPD}(x,t) = f(x,t)+f(x-\pi,t+\pi)$ is the reversal point density (RPD) in the weakly nonlinear and weak diffusion limit as $\epsilon\to 0+$. This limit is also called the {\em weak signaling limit} \cite{igo04}. In (\ref{eq15}), $n(x,t,\phi)$ is continuous and $2\pi$-periodic in $\phi$. The constant solution, $f=\hat{N}/(2\pi)$, is a particular solution of (\ref{eq11}). 

Igoshin et al have derived a Fokker-Planck equation for $f$ in the weak signaling limit as $\epsilon\to 0$ by using physical arguments and also by singular perturbation methods (see Appendix in \cite{igo04}). Their derivation missed the collision-type source term that we find in this paper. Finding this term requires delving more deeply in the perturbation method, therefore we describe this method from scratch. In the limit as $\epsilon\to 0$, we seek a solution 
\begin{eqnarray}
n= n_0(x,t,\phi,\tau)+\epsilon\, n_1(x,t,\phi,\tau) + O(\epsilon^2), \quad\,\tau=\frac{\epsilon t}{2\pi},\label{eq16}\\
n_0(x,t,\phi,\tau)=\left\{\begin{array}{cc}
f(x-\phi,t-\phi,\tau), & 0<\phi<\pi,\\
f(x+\phi,t-\phi,\tau), & -\pi<\phi< 0,\end{array}\right. \label{eq17}
\end{eqnarray}
so that $n_1$ is $2\pi$-periodic in $\phi$. Note that the function $f$ in (\ref{eq17}) has the form (\ref{eq15}) with an additional dependence upon the slow time $\tau$. The equation for $n_1$ is
\begin{eqnarray}
[\partial_t + \mbox{sign}(\phi)\partial_x + \partial_\phi] n_1 = \!\left(\mathcal{D}_x\partial_x^2 +\mathcal{D}_{\phi}\partial_\phi^2-2\pi\Omega_-\chi_{[\alpha,\pi]}\partial_\phi-2\pi\Omega_+\chi_{[-\pi+\alpha,0]}\partial_\phi-\partial_\tau\right)\! \frac{n_0}{2\pi} , \label{eq18}
\end{eqnarray}
with 
\begin{eqnarray}
&&N_+(x,t,\tau)=\int_0^\pi f(x-\psi,t-\psi,\tau)\, d\psi, \nonumber\\
&& N_-(x,t,\tau)=\int_{-\pi}^0 f(x+\psi,t-\psi,\tau)\,d\psi=\int_0^\pi f(x-\psi,t+\psi,\tau)\,d\psi,\nonumber\\
&& \Omega_\pm(x,t,\tau)=\Omega(N_\pm(x,t,\tau)),\quad N_\pm(x,t,\tau)=\int_0^\pi f(x-\psi,t\mp\psi,\tau)\, d\psi. \label{eq19}
\end{eqnarray}
Continuity of the flux \eqref{eq4}, $J=\{1+\epsilon\Omega_-\chi_{[\alpha,\pi]}+\epsilon\Omega_+\chi_{[-\pi+\alpha,0]}\}n-\mathcal{D}_\phi \partial_\phi n/(2\pi)$, across angle boundaries and (\ref{eq16}) yield
\begin{eqnarray}
&& [n_1]_{\phi=0} \equiv n_1|_{\phi=0+}-n_1|_{\phi=0-}= \Omega_+ n_0|_{\phi=0-}+\frac{\mathcal{D}_\phi}{2\pi}[\partial_\phi n_0]|_{\phi=0}, \label{eq20}\\
&& [n_{1}]_{\phi=\alpha} \equiv n_1|_{\phi=\alpha+}-n_1|_{\phi=\alpha-}= -\Omega_- n_0|_{\phi=\alpha+}+\frac{\mathcal{D}_\phi}{2\pi}[\partial_\phi n_0]|_{\phi=\alpha}, \label{eq21}\\
&& [n_1]_{\phi=-\pi+\alpha} \equiv n_1|_{\phi=-\pi+\alpha+}-n_1|_{\phi=-\pi+\alpha-}= -\Omega_+ n_0|_{\phi=-\pi+\alpha+}+\frac{\mathcal{D}_\phi}{2\pi}[\partial_\phi n_0]|_{\phi=-\pi+\alpha}, \label{eq22}\\
&& [n_1]_{\phi=\pi} \equiv n_1|_{\phi=-\pi+}-n_1|_{\phi=\pi-}= \Omega_- n_0|_{\phi=\pi-}+\frac{\mathcal{D}_\phi}{2\pi}[\partial_\phi n_0]|_{\phi=\pi}. \label{eq23}
\end{eqnarray}

From (\ref{eq17}) and (\ref{eq18}), we obtain along the characteristics
\begin{eqnarray}
x(\phi)=x_0+\mbox{sign}(\phi)\phi,\quad t(\phi)=t_0+\phi,\quad f(\phi)=f(x_0,t_0), \label{eq24}
\end{eqnarray}
the following equation for $n_1$:
\begin{eqnarray}
\frac{dn_1}{d\phi}=\left\{\begin{array}{cc}
\{L_+ +\chi_{[\alpha,\pi]}(\partial_{x_0}+\partial_{t_0})\Omega_-(x_0+\phi,t_0+\phi)\}f, & 0<\phi<\pi,\\
\{L_- -\chi_{[\alpha-\pi,0]} (\partial_{x_0}-\partial_{t_0})\Omega_+(x_0-\phi,t_0+\phi)\}f, & -\pi<\phi< 0.\end{array}\right. \label{eq25}
\end{eqnarray}
In (\ref{eq25}) we have defined
\begin{eqnarray}
L_\pm &=& \frac{1}{2\pi}[\mathcal{D}_x\partial_{x_0}^2+\mathcal{D}_\phi(\partial_{x_0}\pm\partial_{t_0})^2-\partial_\tau]. \label{eq26}
\end{eqnarray}

Ignoring the initial condition for $n_1$, the solution of (\ref{eq25}) along the characteristics (\ref{eq24}) is
\begin{eqnarray}
n_1^p=\left\{\begin{array}{cc}
\phi L_+f+\left(1-\frac{\alpha}{\pi}\right)\Omega_-(x_0+\alpha,t_0+\alpha,\tau)\, f, & 0<\phi<\alpha\\ 
\phi L_+f+ (\partial_{x_0}+\partial_{t_0})f\int_\alpha^\phi\Omega_- d\phi+C_\alpha, & \alpha<\phi<\pi,\\
\phi L_- f +(\partial_{x_0}-\partial_{t_0})f\int_\phi^0\Omega_+ d\phi+ C_0, & \alpha-\pi<\phi<0,\\ 
\phi L_- f+ C_{\alpha-\pi}, & -\pi<\phi<\alpha-\pi,\end{array}\right. \label{eq27}
\end{eqnarray}
where $f=f(x_0,t_0,\tau)$, the constants of integration $C_j$ are independent of $\phi$, and
\begin{eqnarray}
\left\{\begin{array}{c}
\int_\alpha^\phi\Omega_- d\phi=\int_\alpha^\phi\Omega\left(\int_0^\pi f(x_0+\phi'-\psi,t_0+\phi'+\psi,\tau)d\psi\right) d\phi',\\
\int_\phi^0\Omega_+ d\phi=\int_\phi^0\Omega\left(\int_0^\pi f(x_0-\phi'-\psi,t_0+\phi'-\psi,\tau)d\psi\right) d\phi'.\end{array}\right. \label{eq28}
\end{eqnarray}
To determine the constants $C_j$ in \eqref{eq27}, we impose the jump conditions (\ref{eq20})-(\ref{eq22}). Using \eqref{eq17}, $[\partial_\phi n_0]|_{\phi=\alpha}=[\partial_\phi n_0]|_{\phi=-\pi+\alpha}=0$, $[\partial_\phi n_0]|_{\phi=\pi}=2\partial_xf(x_0-\pi,t_0+\pi,\tau)$, $[\partial_\phi n_0]|_{\phi=0}=-2\partial_xf(x_0,t_0,\tau)$, and we find 
\begin{eqnarray}
C_\alpha&=& -\frac{\alpha}{\pi}\,\Omega_-(x_0+\alpha,t_0+\alpha,\tau)\, f(x_0,t_0,\tau),
\label{eq29}\\
C_0&=& \left[\left(1-\frac{\alpha}{\pi}\right)\!\Omega_-(x_0+\alpha,t_0+\alpha,\tau)-\Omega_+(x_0,t_0)\right] f(x_0,t_0,\tau)+\frac{\mathcal{D}_\phi}{\pi}\partial_x f(x_0,t_0,\tau), 
\label{eq30}\\
C_{\alpha-\pi}&=& \left[\Omega_+(x_0-\alpha+\pi,t_0+\alpha-\pi,\tau)+\!\left(1-\frac{\alpha}{\pi}\right)\!\Omega_-(x_0+\alpha,t_0+\alpha,\tau)-\Omega_+(x_0,t_0)\right]\nonumber\\
&\times&
f(x_0,t_0,\tau) + (\partial_{x_0}-\partial_{t_0})\left(f(x_0,t_0,\tau)\int_{\alpha-\pi}^0\Omega_+d\phi\right)\!+\frac{\mathcal{D}_\phi}{\pi}\partial_x f(x_0,t_0,\tau). \label{eq31}
\end{eqnarray}
The integral on the right hand side of \eqref{eq31} can be rewritten as 
\begin{eqnarray}
\int_{\alpha-\pi}^0\Omega_+d\phi&=& \int_{\alpha-\pi}^0\Omega\left(\int_0^\pi f(x_0-\phi-\psi,t_0+\phi-\psi,\tau)d\psi\right)d\phi\nonumber\\
&=& \int_{\alpha}^\pi \Omega\left(\int_0^\pi f(x_0+\pi-\phi-\psi,t_0-\pi+\phi-\psi,\tau)d\psi\right)d\phi\nonumber\\
&=& \int_{\alpha}^\pi \Omega\left(\int_0^\pi f(x_0+\pi-\phi-\psi,t_0+\pi+\phi-\psi,\tau)d\psi\right)d\phi\nonumber\\
&=& \int_{\alpha}^\pi \Omega\left(\int_0^\pi f(x_0-\phi+\psi,t_0+\phi+\psi,\tau)d\psi\right)d\phi. \label{eq32}
\end{eqnarray}
We have used that $f(x,t,\tau)$ is $2\pi$-periodic in $t$ and the change of variable $\pi-\psi\to\psi$ to simplify the integral in (\ref{eq32}). The first term in $C_{\alpha-\pi}$ can be similarly simplified, thereby producing
\begin{eqnarray}
C_{\alpha-\pi}&=& \left[\Omega\left(\int_0^{\pi} f(x_0-\alpha+\psi,t_0+\alpha+\psi,\tau)d\psi\right)-\Omega\left(\int_0^{\pi} f(x_0-\psi,t_0-\psi,\tau)d\psi\right)\right.\nonumber\\ 
&+&\left.\left(1-\frac{\alpha}{\pi}\right)\Omega\left(\int_0^{\pi} f(x_0+\alpha-\psi,t_0+\alpha+\psi,\tau)d\psi\right)  \right]f \nonumber\\
&+& (\partial_{x_0}-\partial_{t_0})\left[f\int_{\alpha}^\pi \Omega\left(\int_0^\pi f(x_0-\phi+\psi,t_0+\phi+\psi,\tau)d\psi\right)d\phi\right]\!+\frac{\mathcal{D}_\phi}{\pi}\partial_x f.\label{eq33}
\end{eqnarray}

The condition (\ref{eq23}) ensuring $2\pi$-periodicity in $\phi$ provides the sought equation for $f$:
\begin{eqnarray}
&&\partial_\tau f + \partial_x (Uf)+\partial_t(Vf)- (\mathcal{D}_x+\mathcal{D}_\phi)\partial_x^2f -\mathcal{D}_\phi \partial_t^2 f= f\, Q[f]\nonumber\\
&&\quad\quad-\frac{\mathcal{D}_\phi}{\pi}\partial_x[f(x,t,\tau)-f(x-\pi,t+\pi,\tau)],\label{eq34}\\
&&U=\int_{\alpha}^\pi \left[ \Omega\left(\int_0^\pi f(x-\phi+\psi,t+\phi+\psi,\tau)d\psi\right)\right.\nonumber\\
&&\quad\quad-\left. \Omega\left(\int_0^\pi f(x+\phi-\psi,t+\phi+\psi,\tau)d\psi\right)\right]d\phi, \label{eq35}\\
&&V=-\int_{\alpha}^\pi \left[ \Omega\left(\int_0^\pi f(x-\phi+\psi,t+\phi+\psi,\tau)d\psi\right)\right.\nonumber\\
&&\quad\quad+\left. \Omega\left(\int_0^\pi f(x+\phi-\psi,t+\phi+\psi,\tau)d\psi\right)\right]d\phi,\label{eq36}\end{eqnarray}
\begin{eqnarray}
&&Q[f]= \Omega\left(\int_0^\pi f(x+\psi,t-\psi,\tau)d\psi\right) + \Omega\left(\int_0^\pi f(x-\psi,t-\psi,\tau)d\psi\right)\nonumber\\
&&\quad\quad\quad-\Omega\left(\int_0^\pi f(x+\alpha-\psi,t+\alpha+\psi,\tau)d\psi\right)\nonumber\\
&&\quad\quad\quad-\Omega\left(\int_0^\pi f(x-\alpha+\psi,t+\alpha+\psi,\tau)d\psi\right) .\label{eq37}
\end{eqnarray}
We have dropped the subscripts 0 in the variables $x$ and $t$. Unimportant changes in the notation aside (our $\tau$ corresponds to Igoshin et al's variable $T$, $\mathcal{D}_x=2\pi D_1$, $\mathcal{D}_\phi=2\pi D_2$), these equations are different from those derived by Igoshin et al \cite{igo04}: $U$ and $V$ in (\ref{eq34}) are the same but Igoshin et al's Fokker-Planck equation lacks the source term $fQ[f]$. The reason is that Igoshin et al do not impose consistently the jump conditions (\ref{eq20})-(\ref{eq23}) in their derivation; see their equations (A10)-(A.14) in the Appendix of Ref. \onlinecite{igo04}. The terms in \eqref{eq37} do not appear in these equations.

Equation \eqref{eq34} is a Fokker-Planck type equation with a source term $fQ[f]-\frac{\mathcal{D}_\phi}{\pi}\partial_x[f(x,t,\tau)-f(x-\pi,t+\pi,\tau)]$. Even in the absence of noise ($\mathcal{D}_x=\mathcal{D}_\phi=0$), $fQ[f]$ acts as an effective {\em collision term} that produces dissipation. As we shall see in the next section, the source term provides a mechanism for wave number and speed selection of the ripples. No such mechanism was found in Ref. \onlinecite{igo04}.

\section{Constant solution in the weak signaling limit and its linear stability}
\label{sec:4}
The constant function $f=\hat{N}/(2\pi)$ is an exact solution of (\ref{eq34}) that coincides with the following exact piecewise constant solution of the full model \eqref{eq11}: 
\begin{eqnarray}
&& n_s(\phi)=p\, [\chi_{[0,\alpha]}(\phi)+\chi_{[-\pi,-\pi+\alpha]}(\phi)]+q\, [\chi_{[-\pi+\alpha,0]}(\phi)+\chi_{[\alpha,\pi]}(\phi)], \label{eq38}\\
&&p=\frac{1+\epsilon\,\Omega\left(\frac{\hat{N}}{2}\right)}{\pi+\epsilon\alpha\,\Omega\left(\frac{\hat{N}}{2}\right)}\,\frac{\hat{N}}{2},\quad q=\frac{\frac{\hat{N}}{2}}{\pi+\epsilon\alpha\,\Omega\left(\frac{\hat{N}}{2}\right)} \label{eq39}
\end{eqnarray}
when $\epsilon=0$. In (\ref{eq38}), $p$ and $q$ given by (\ref{eq39}) have been calculated from $N_\pm=\alpha p+(\pi-\alpha)q=\hat{N}/2$ and from the condition that the flux $J=n[1+\epsilon \Omega(N_-)\chi_{[\alpha,\pi]}+\epsilon \Omega(N_+)\chi_{[-\pi+\alpha,0]})]$ should be continuous. Note that substitution of $f=\hat{N}/(2\pi)$ in (\ref{eq27}) yields
\begin{eqnarray}
n_1^p=\left\{\begin{array}{cc}
\left(1-\frac{\alpha}{\pi}\right)\Omega\left(\frac{\hat{N}}{2}\right)\frac{\hat{N}}{2\pi}, & 0<\phi<\alpha\\ 
-\frac{\alpha}{\pi}\,\Omega\left(\frac{\hat{N}}{2}\right)\frac{\hat{N}}{2\pi}, & \alpha<\phi<\pi,\\
-\frac{\alpha}{\pi}\,\Omega\left(\frac{\hat{N}}{2}\right)\frac{\hat{N}}{2\pi}, & \alpha-\pi<\phi<0,\\ 
\left(1-\frac{\alpha}{\pi}\right)\Omega\left(\frac{\hat{N}}{2}\right)\frac{\hat{N}}{2\pi}, & -\pi<\phi<\alpha-\pi,\end{array}\right. \label{eq40}
\end{eqnarray}
after using (\ref{eq29})-(\ref{eq31}). Equation \eqref{eq40} agrees with \eqref{eq38}-\eqref{eq39} up to terms of order $\epsilon^2$.

Let us see what we find by a linear stability analysis of $f=\hat{N}/(2\pi)$ as a solution of (\ref{eq34}). Inserting $f=\hat{N}/(2\pi)+\nu(x,t,\tau)$ with $\nu\ll 1$ in (\ref{eq34}) and keeping only terms that are linear in $\nu$, we find
\begin{eqnarray}
&&\left[\partial_\tau -2(\pi-\alpha)\, \Omega\left(\frac{\hat{N}}{2}\right)\partial_t- (\mathcal{D}_x+\mathcal{D}_\phi)\partial_x^2 -\mathcal{D}_\phi \partial_t^2\right]\!\nu=\frac{\hat{N}}{2\pi}\,\Omega'\left(\frac{\hat{N}}{2}\right)\nonumber\\
&&\times\left\{\partial_t\int_{\alpha}^\pi\int_0^\pi [\nu(x-\phi+\psi,t+\phi+\psi,\tau)+\nu(x+\phi-\psi,t+\phi+\psi,\tau)]d\psi d\phi\right.\nonumber\\
&&\quad- \partial_x\int_{\alpha}^\pi\int_0^\pi [\nu(x-\phi+\psi,t+\phi+\psi,\tau)-\nu(x+\phi-\psi,t+\phi+\psi,\tau)]d\psi d\phi \nonumber\\
&&\quad+\int_0^\pi [\nu(x+\psi,t-\psi,\tau) + \nu(x-\psi,t-\psi,\tau)-\nu(x+\alpha-\psi,t+\alpha+\psi,\tau)\nonumber\\
&&\quad\quad\quad- \left.\nu(x-\alpha+\psi,t+\alpha+\psi,\tau)]\, d\psi\right\} -\frac{\mathcal{D}_\phi}{\pi}\partial_x[\nu(x,t,\tau)-\nu(x-\pi,t+\pi,\tau)].
\label{eq41}
\end{eqnarray} 

Assuming $\nu=e^{ikx+ilt+\sigma\tau}$, we obtain the following eigenvalues: 
\begin{eqnarray}
\sigma&=& 2il(\pi-\alpha)\, \Omega\left(\frac{\hat{N}}{2}\right)- (\mathcal{D}_x+\mathcal{D}_\phi)k^2 -\mathcal{D}_\phi l^2 + i2\hat{N}\,\Omega'\left(\frac{\hat{N}}{2}\right)\nonumber\\
&\times&\left\{e^{i\pi l}\left[e^{i\alpha L}\sin[(\pi-\alpha)L]\frac{\sin(\pi K)}{2 \pi K}+e^{i\alpha K}\sin[(\pi-\alpha)K]\frac{\sin(\pi L)}{2 \pi L}\right]\right.\nonumber\\
&-&\left. e^{i\alpha L}\sin(\alpha L+\pi K)\frac{\sin(\pi K)}{2\pi K} -e^{i\alpha K}\sin(\alpha K+\pi L)\frac{\sin(\pi L)}{2\pi L}\right\}\!-\frac{ik\mathcal{D}_\phi}{\pi}(1-e^{i2\pi L}) ,
\label{eq42}
\end{eqnarray} 
where
\begin{eqnarray}
K=\frac{l+k}{2},\quad L= \frac{l-k}{2}. \label{eq43}
\end{eqnarray} 
For real $k$ and $l$, the real and imaginary parts of (\ref{eq42}) are
\begin{eqnarray}
\mbox{Re }\sigma&=& - (\mathcal{D}_x+\mathcal{D}_\phi)k^2 -\mathcal{D}_\phi l^2 - \hat{N}\,\Omega'\left(\frac{\hat{N}}{2}\right)\left\{\sin(\alpha L+\pi l)\sin[(\pi-\alpha)L]\frac{\sin(\pi K)}{\pi K}\right.\nonumber\\
&+&\sin(\alpha K+\pi l)\sin[(\pi-\alpha)K]\frac{\sin(\pi L)}{\pi L}-\sin(\alpha L)\sin(\alpha L+\pi K)\frac{\sin(\pi K)}{\pi K}\nonumber\\
&-&\left. \sin(\alpha K)\sin(\alpha K+\pi L)\frac{\sin(\pi L)}{\pi L}\right\}\!-\frac{k\mathcal{D}_\phi}{\pi}\sin (2\pi L) ,
\label{eq44}\\
\mbox{Im }\sigma&=& 2l(\pi-\alpha)\,\Omega\left(\frac{\hat{N}}{2}\right) + \hat{N}\,\Omega'\left(\frac{\hat{N}}{2}\right)\left\{\cos(\alpha L+\pi l)\sin[(\pi-\alpha)L]\frac{\sin(\pi K)}{\pi K}\right.\nonumber\\
&+& \cos(\alpha K+\pi l)\sin[(\pi-\alpha)K]\frac{\sin(\pi L)}{\pi L}-\cos(\alpha L)\sin(\alpha L+\pi K)\frac{\sin(\pi K)}{\pi K}\nonumber\\
&-&\left. \cos(\alpha K)\sin(\alpha K+\pi L)\frac{\sin(\pi L)}{\pi L}\right\}\!-\frac{2k\mathcal{D}_\phi}{\pi}\sin^2(\pi L) ,
\label{eq45}
\end{eqnarray} 
respectively. Except for the last three terms in (\ref{eq44}), this is the same as (57) in Ref. \onlinecite{igo04} (with $\hat{N}$ replaced by $\hat{N}/2$) provided $l$ is an integer. We have $\sigma=0$ for $k=l=0$ indicating that we can shift the constant solution $f=\hat{N}/(2\pi)$ by an arbitrary quantity. The integral condition (\ref{eq12}) fixes the value of $\hat{N}$ and therefore we have to ignore the zero eigenvalue.

\begin{figure}[ht]
\begin{center}
\includegraphics[width=11cm]{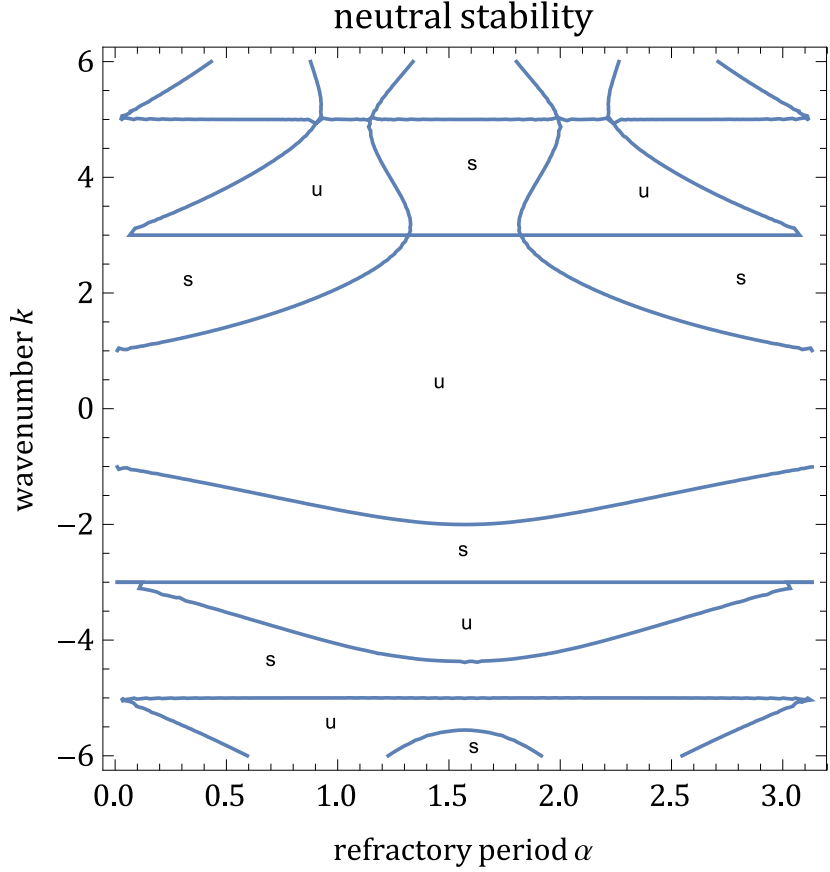} \qquad
\end{center}
\caption{Contour plot of the neutral stability curve Re\,$\sigma(\alpha,k,1)=0$ for $0\leq\alpha\leq\pi$,  $-6\leq k\leq 6$ and no diffusion, $\mathcal{D}_x=\mathcal{D}_\phi=0$. Patterns are expected in regions where the constant solution is unstable (marked with ``u'') and they may disappear leaving the constant solution in regions where the latter is linearly stable (marked with ``s'').
\label{fig1}}
\end{figure}

Note that as $r\to\infty$, $\Omega'(\hat{N}/2)\to 0$ and the constant solution becomes stable according to (\ref{eq44}). At bifurcation points, Re $\sigma=0$, Im $\sigma\neq 0$. This suggests that a Hopf bifurcation occurs, which our numerical simulations support. Pattern solutions that are periodic in the slow time issue forth from the constant solution as {\em supercritical} Hopf bifurcations according to our numerical evidence. We have not found examples of subcritical Hopf bifurcations and hysteresis.  

As $\epsilon\to 0$, $f(x,t,\tau)$ approaches \eqref{eq17}, which is a $2\pi$-periodic function of $t$. The corresponding frequency is $l=1$. Figure \ref{fig1} shows the contour plot of the neutral stability curve Re $\sigma(\alpha,k,1)=0$ in \eqref{eq44} for frequency $l=1$ in the absence of diffusion, $\mathcal{D}_x=\mathcal{D}_\phi=0$. The real part of the eigenvalue \eqref{eq44}, Re $\sigma(\alpha,k,1)$, is positive at the middle region enclosing $k=0$ in Fig. \ref{fig1} and its sign changes each time a line of the neutral stability curve is crossed. Thus, for unit frequency and $|k|\leq 1$, we expect to see patterns in $x$ and $t$ provided there is no diffusion. For any value of the refractory period $\alpha$, the constant solution is unstable for waves traveling to the left ($k=-1$) and also for waves traveling to the right ($k=1$). Let us assume that the initial condition is a periodic pattern of wave number $k>0$. Increasing the wave number $k$ or adding diffusion (which may be the result of unavoidable numerical errors) stabilize the constant solution and cause the patterns to disappear. For a fixed value of the refractory period $\alpha$, the neutral stability curve of Fig.~\ref{fig1} yields the critical wave number below which patterns with that wave number appear. Similarly, increasing the refractory period from $\alpha=0$ at a fixed wave number $1<k<3$ should produce patterns with wave number $k$ once $\alpha$ surpasses the critical value given by the neutral stability curve.

For other values of the frequency, the neutral stability curve qualitatively changes. For instance, let us consider wave trains traveling to the right, so that $k=-l$, $L=l$, $K=0$, as in \cite{igo04}. (\ref{eq44}) becomes
\begin{eqnarray}
\mbox{Re }\sigma\!&=& - (\mathcal{D}_x+2\mathcal{D}_\phi) l^2 - \hat{N}\,\Omega'\left(\frac{\hat{N}}{2}\right)\left\{\sin[(\alpha+\pi) l]\sin[(\pi-\alpha)l]-\sin^2(\alpha l)\right\}\!+\frac{l\mathcal{D}_\phi}{\pi}\sin (2\pi l)\nonumber\\
&=& - \mathcal{D}_x l^2 - \frac{l\mathcal{D}_\phi}{\pi}[2\pi l-\sin (2\pi l)] + \hat{N}\,\Omega'\left(\frac{\hat{N}}{2}\right)[2\sin^2(\alpha l)-\sin^2(\pi l)] ,  \label{eq46}\\
\mbox{Im }\sigma\!&=&\! 2l(\pi-\alpha)\Omega\!\left(\frac{\hat{N}}{2}\right)\! + \hat{N}\Omega'\!\left(\frac{\hat{N}}{2}\right)\!\left\{\cos[(\pi+\alpha) l]\sin[(\pi-\alpha)l]-\frac{1}{2}\sin(2\alpha l)\right.\nonumber\\
&-&\left.\frac{\sin^2(\pi l)}{\pi l}\right\}\!+\frac{2l\mathcal{D}_\phi}{\pi}\sin^2(\pi l)= 2l(\pi-\alpha)\,\Omega\!\left(\frac{\hat{N}}{2}\right)\! + \frac{\hat{N}}{2}\,\Omega'\!\left(\frac{\hat{N}}{2}\right)\!\left[\sin(2\pi l)\right.\nonumber\\
&-&\left. \frac{2\sin^2(\pi l)}{\pi l}-2\sin(2\alpha l)\right]\!+ \frac{2l\mathcal{D}_\phi}{\pi}\sin^2(\pi l). \label{eq47}
\end{eqnarray} 

\begin{figure}[ht]
\begin{center}
\includegraphics[width=11cm]{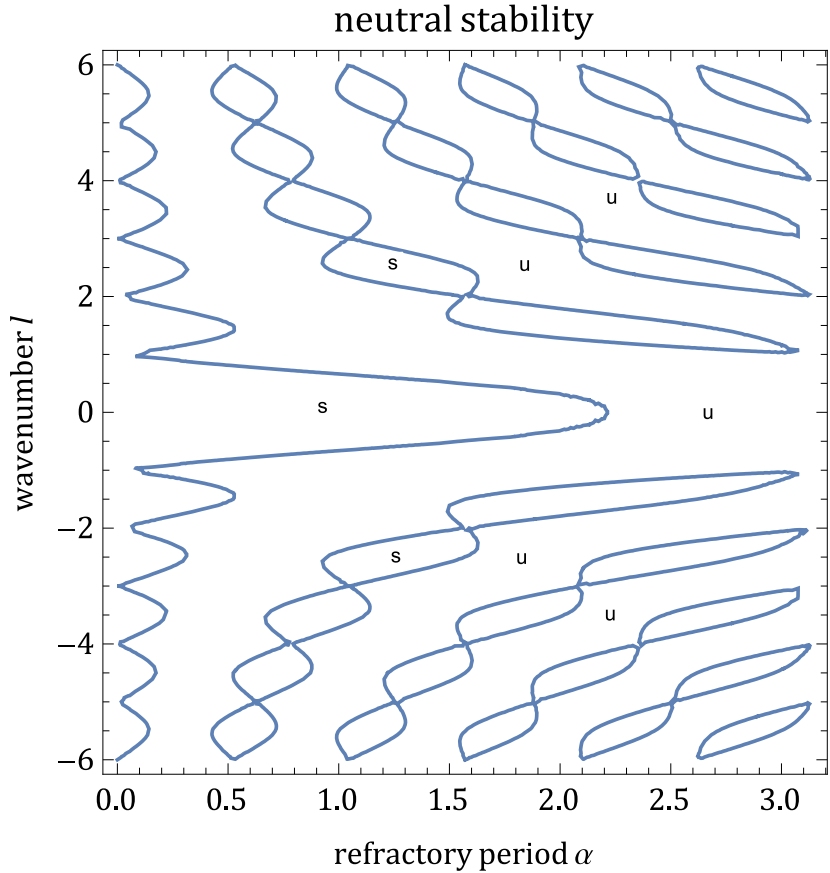} \qquad
\end{center}
\caption{Contour plot of the neutral stability curve Re\,$\sigma(\alpha,-l,l)=0$ (waves traveling to the right) for $0\leq\alpha\leq\pi$, $-6\leq l\leq 6$, and $\mathcal{D}_x=\mathcal{D}_\phi=0$. Meaning of ``s'' and ``u'' as in Fig.~\ref{fig1}.
\label{fig2}}
\end{figure}

In absence of diffusion, $\mathcal{D}_x=\mathcal{D}_\phi=0$, \eqref{eq46} shows that Re $\sigma=0$ for $l = l_c(\alpha)$ such that $2\sin^2(\alpha l)=\sin^2(\pi l)$. This $l_c(\alpha)$ provides the wavelength of the observed patterns. The contour plot of Re $\sigma=0$ in Fig. \ref{fig2} shows that the function $l_c(\alpha)$ is multivalued and that its leftmost branch (corresponding to the lowest values of $l>0$) decreases from $\alpha=\pi/\sqrt{2}$ at $l=0$ to 0 at $l=1$. For this parameter range, the constant solution is linearly stable if $l<l_c(\alpha)$ and unstable for $l>l_c(\alpha)$ (but smaller than values at the next branch of $l_c(\alpha)$). For sufficiently small $l$, $\alpha$ smaller (larger) than a critical value corresponds to stable (unstable) constant solution. As the exponent $r$ increases and $\Omega'(\hat{N}/2)$ correspondingly decreases, the contribution of the second term in (\ref{eq46}) becomes smaller and can be obliterated by diffusion or numerical noise even if the said term is positive.

\section{Numerical solutions}
\label{sec:5}
In this section, we present numerical solutions of the full kinetic model equations (\ref{eq1}) - (\ref{eq6}) and compare them with the linear stability results for the Fokker-Planck type equation (\ref{eq34}) derived in the weak signaling limit. To visualize the results, we obtain the total density, $N_+(x,t)+N_-(x,t)$, and the RPD, $n_{\rm RPD}(x,t)$, from the numerical solution. 

We construct the numerical solution using uniform grids for both the spatial variable $x$ and the phase variable $\phi$ extended over their whole domains and assigning an approximate value of the solution on every point of the two-dimensional grid at every time step. In all our calculations in this paper we use $N=100$ subintervals in both space and phase and set ${\cal L}=\pi$, $r=4$. Then we solve the noiseless nondimensional equations, \eqref{eq11} (with $\mathcal{D}_x=\mathcal{D}_\phi=0$), using as basic scheme an upwind finite difference scheme in conservation form for the spatial terms (space and phase) and the Euler explicit algorithm to evolve in time. The computation of the integral form of the fluxes for the internal clock variable is performed through the trapezoidal rule of numerical integration extended over the whole domain of the phase variable, using the approximated values of the solution at the grid points. Explicit upwind schemes require a Courant-Friedrichs-Lewy (CFL) restriction on the time step, $\Delta t$, in terms of the maximum wave speed and the spatial stepsizes $\Delta x$ and $\Delta \phi$, of the form:
\begin{equation}
\frac{\Delta t}{(\Delta x^2 + \Delta \phi^2)^{1/2}}\leq \frac{C}{ 1 +(1+\epsilon) }.\label{eq48}
\end{equation}
Here $C$ is positive constant such that $C<1$ to ensure stability of the scheme. We use $C=0.4$ in all our computations. 

Since the first order scheme introduces more dissipation than needed, we have formulated a high order accurate version in space and time based on the fifth-order weighted essentially nonoscillatory reconstruction procedure (WENO5) for the spatial variables and a third-order Runge-Kutta method to evolve in time. WENO5 procedure was designed to get fifth order accuracy in space using a nonlinear convex combination of essentially nonoscillatory parabolas, (see \cite{jia96} for details). The third order Runge-Kutta method for the integration in time is the one proposed by Shu and Osher in \cite{shuosher89} and allows the maximum time stepsize dictated by the first order upwind scheme.

If $\epsilon=0$ the hyperbolic system (\ref{eq1})-\eqref{eq6} becomes a system of linear wave equations and the numerical solution can be approximated using standard upwind schemes of any order of accuracy in space and time. When nonlinearity is present, $\epsilon>0$, the solution might present  wave steepening but the non-locality of the flux in Eq. (\ref{eq3}) introduces an analytic dissipation mechanism. Then even for zero-diffusion, the solution is well defined and plain upwind schemes converge stably to it. The jump conditions (\ref{eq20})-(\ref{eq23}) may enforce continuity at the boundaries $\phi=\pm\pi$, 0, $\alpha$, $\alpha-\pi$, but wave steepening can be generated in the  spatial direction of the solution for short periods of time. On the other hand, we do not have numerical evidence of shock wave formation. The dissipation mechanism introduced by the non-local fluxes might prevent the formation of shock waves. This issue will be examined in the near future.

\begin{figure}[ht]
\begin{center}
\includegraphics[width=8cm]{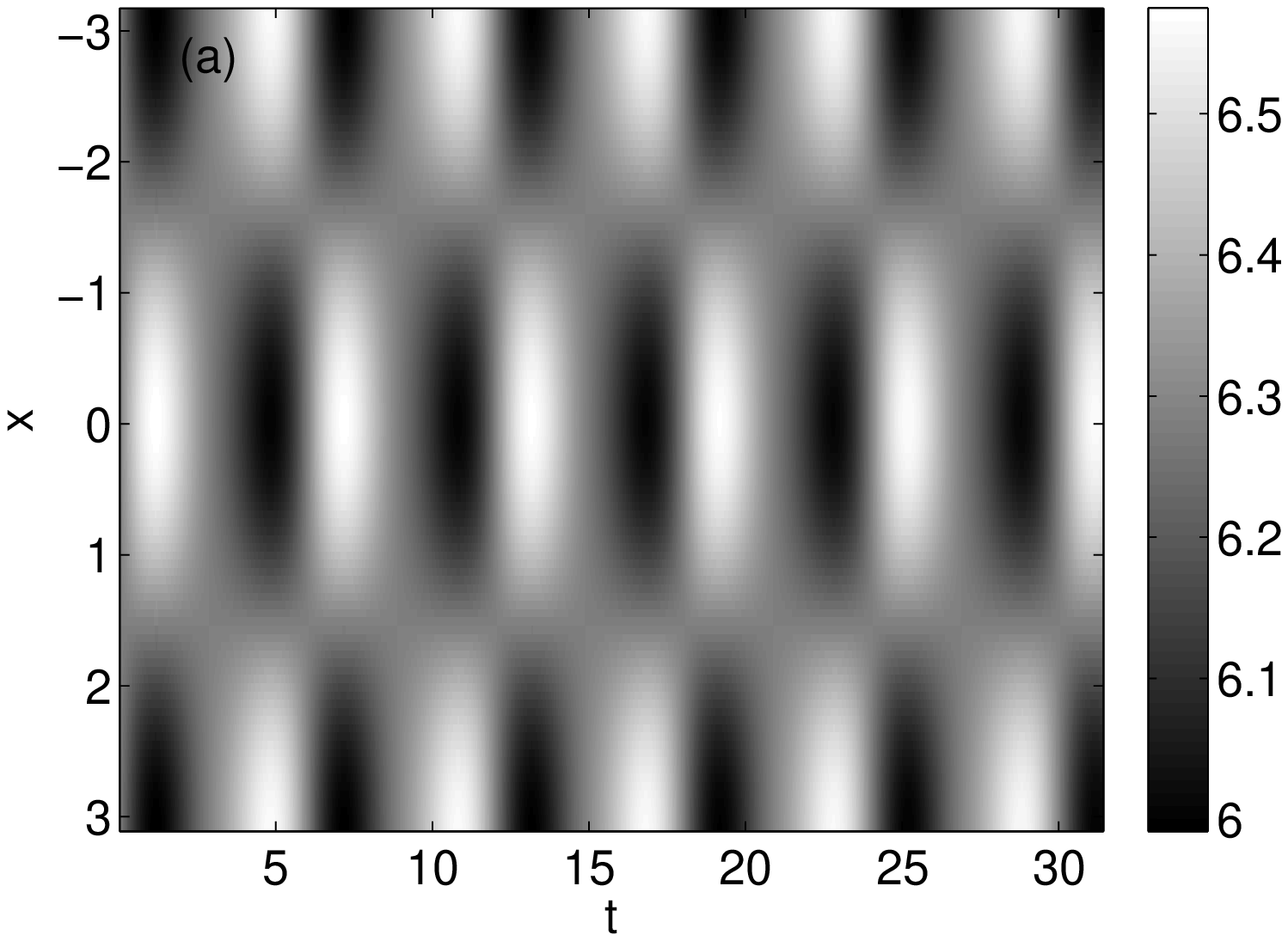} %\qquad
\includegraphics[width=8cm]{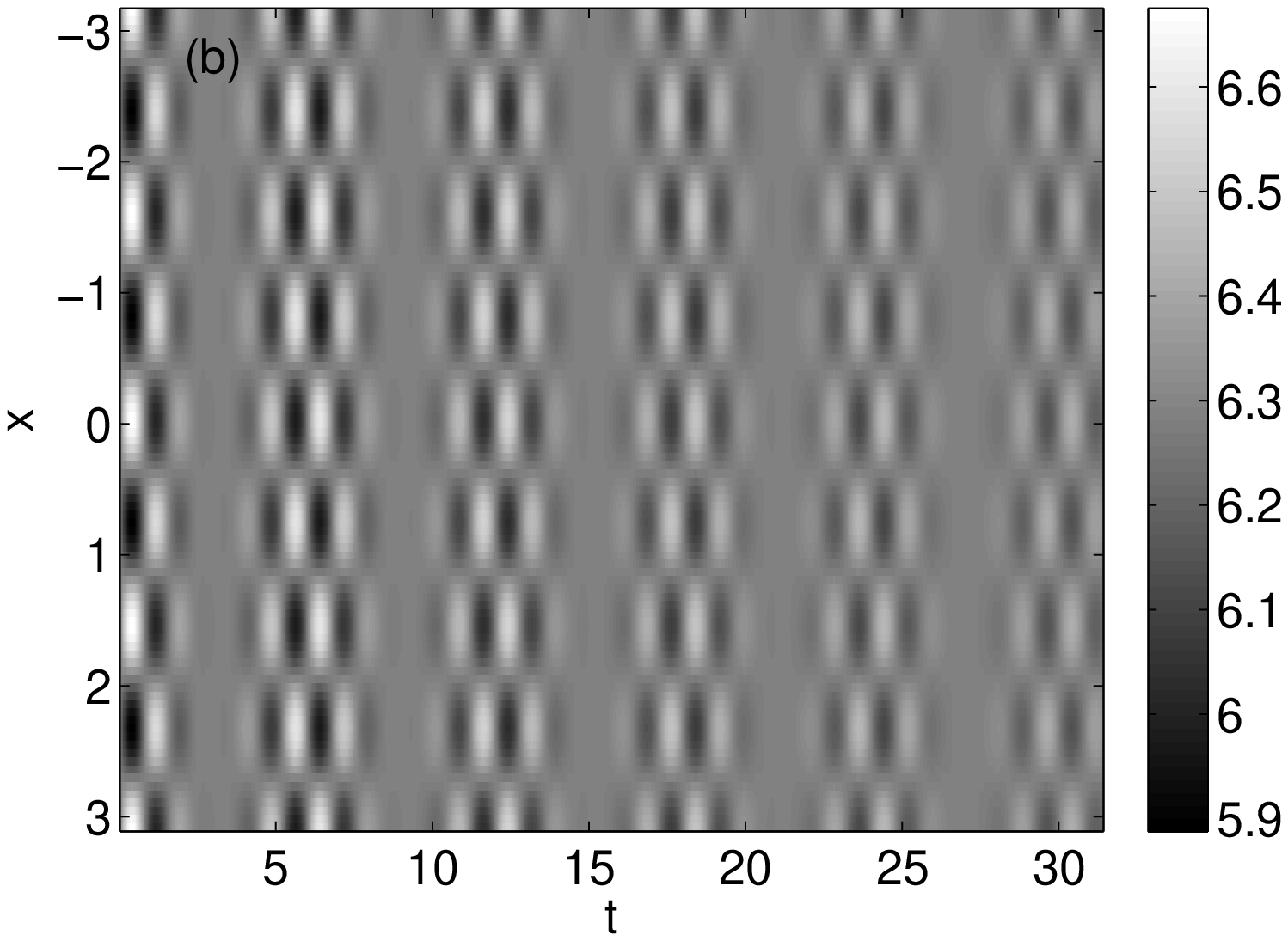} 
\includegraphics[width=8cm]{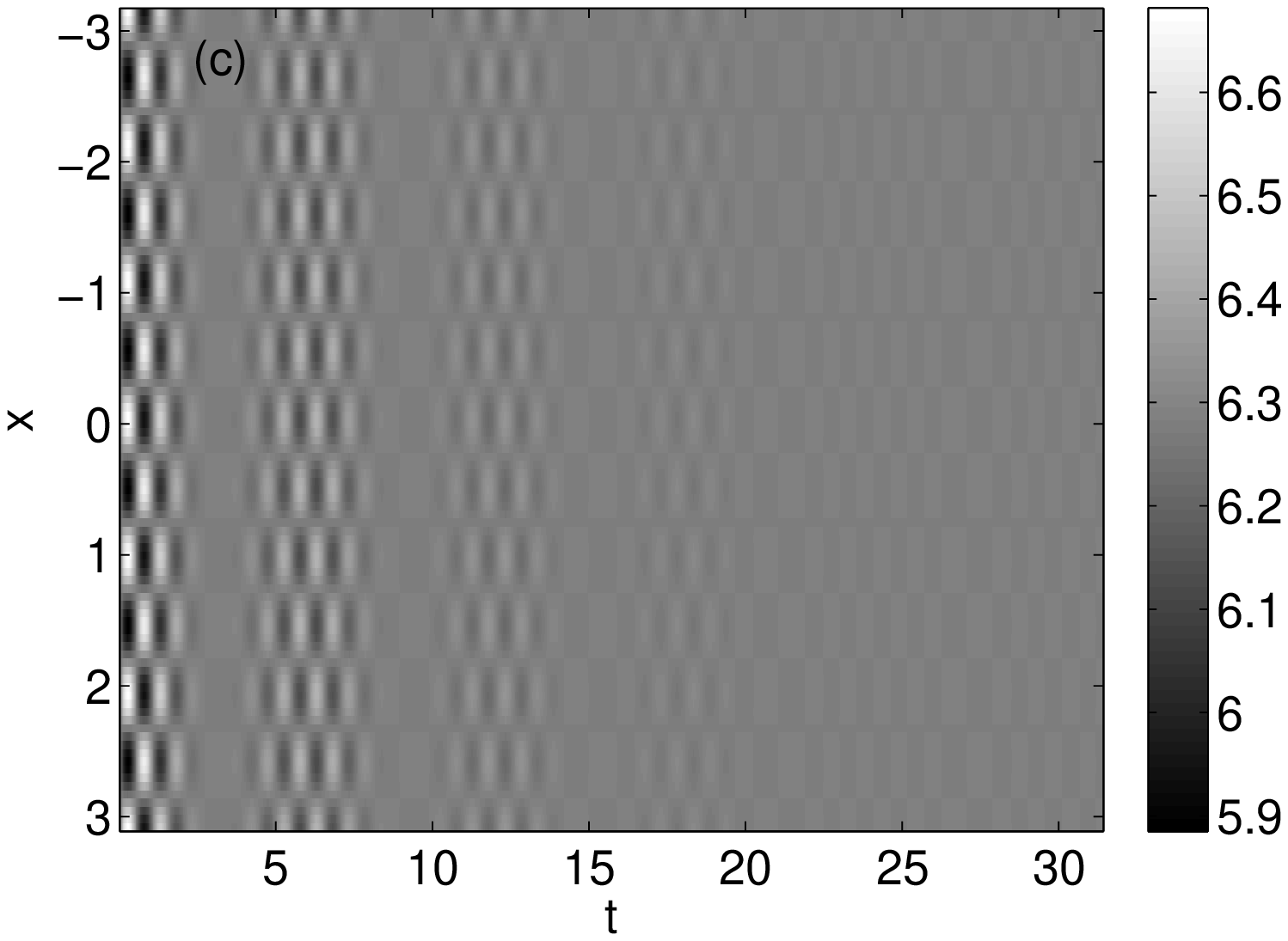} 
\end{center}
\caption{Contour plots of the total density $N_+(x,t)+N_-(x,t)$ for $\alpha=\pi/2$,  $\epsilon=0.1$ and wavenumber: (a) $k=1$, (b) $k=4$, (c) $k=6$.  
\label{fig3}}
\end{figure}

Igoshin et al \cite{igo01} use a heuristic argument to suggest that $k_c=1$ and estimate $\epsilon=3$, $\alpha=\pi/5$ from their measurements. Different values of $\epsilon$ give idea of the solution either close or far from the weak signaling limit: $\epsilon=0.1$ corresponds to the weak signaling limit and larger values, such as $\epsilon=3$ \cite{igo01}, or $\epsilon=10$, go beyond the Fokker-Planck type description of Section \ref{sec:4}. To check the stability results of that section, we first solve the nondimensional equation \eqref{eq11} with the following initial condition that is periodic in $x$ and $\phi$: $n(x,\phi,t=0)=1+0.1 \sin (kx)\sin (\phi )$ on the interval $[-\pi ,\pi ]\times \lbrack -\pi ,\pi ]$. For a first numerical simulation, we consider the case of $\epsilon=0.1$ and a fixed  $\alpha =\pi /2$. In Figure~\ref{fig3}, we show the time-space surface plots of the total density with $k=1$, $k=4$, and $k=6$ from left to right. As expected from the neutral stability curve of Fig.~\ref{fig1}, the simulations of Fig.~\ref{fig3} show that periodic patterns with angular frequency $l=1$ are found for $k=1$, whereas the uniform stationary solution is linearly stable if $k=4$. Fig.~\ref{fig3}(b) shows a transient stage towards this solution. At $k=6$, the uniform stationary solution is unstable to periodic patterns. However, simulations show an even faster transient to the uniform stationary solution. According to \eqref{eq44}, noise has a strongly stabilizing effect of this solution at high wavenumber so that numerical noise might have obliterated the pattern solution at $k=6$. 

\begin{figure}[ht]
\begin{center}
\includegraphics[width=8cm]{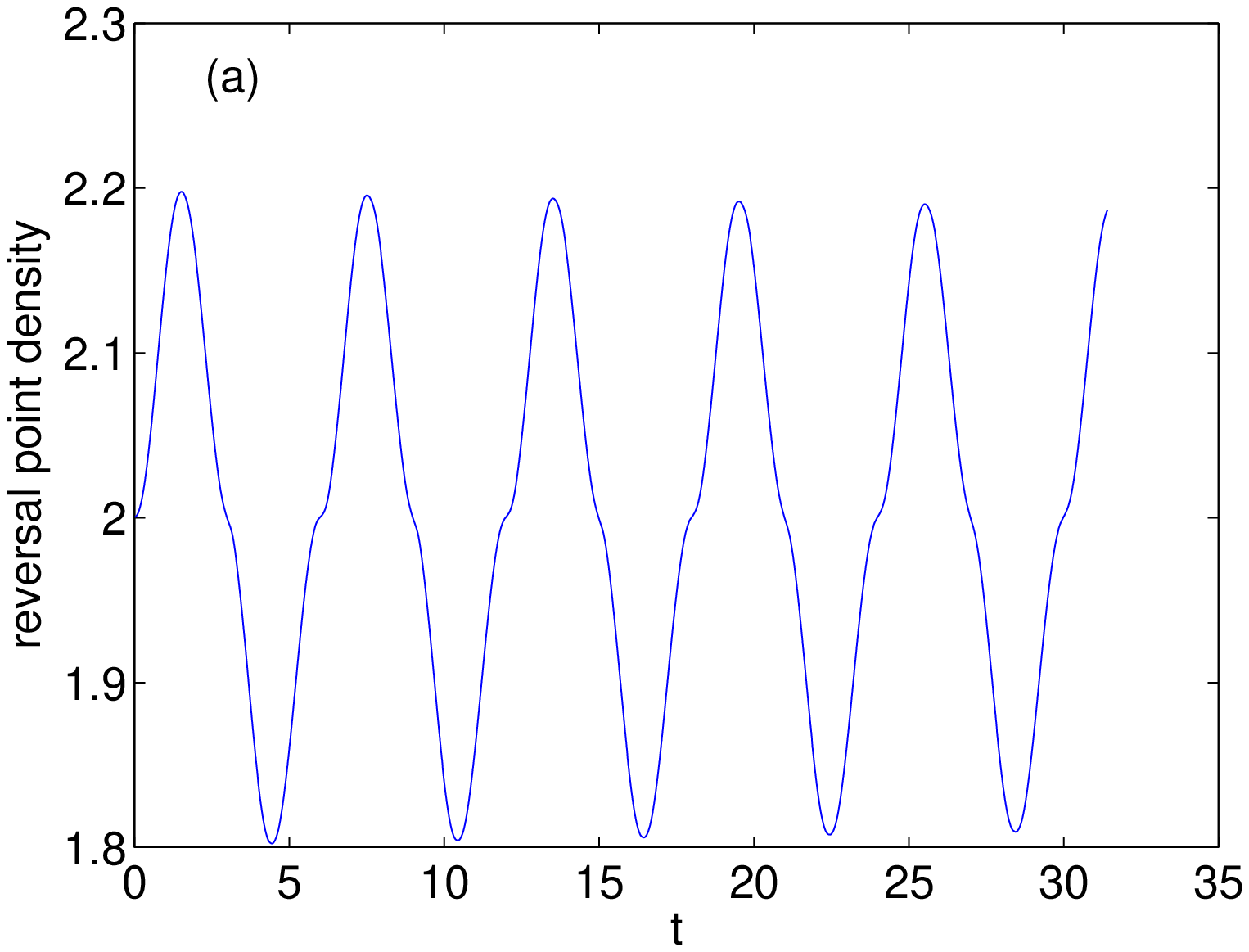} %\qquad
\includegraphics[width=8cm]{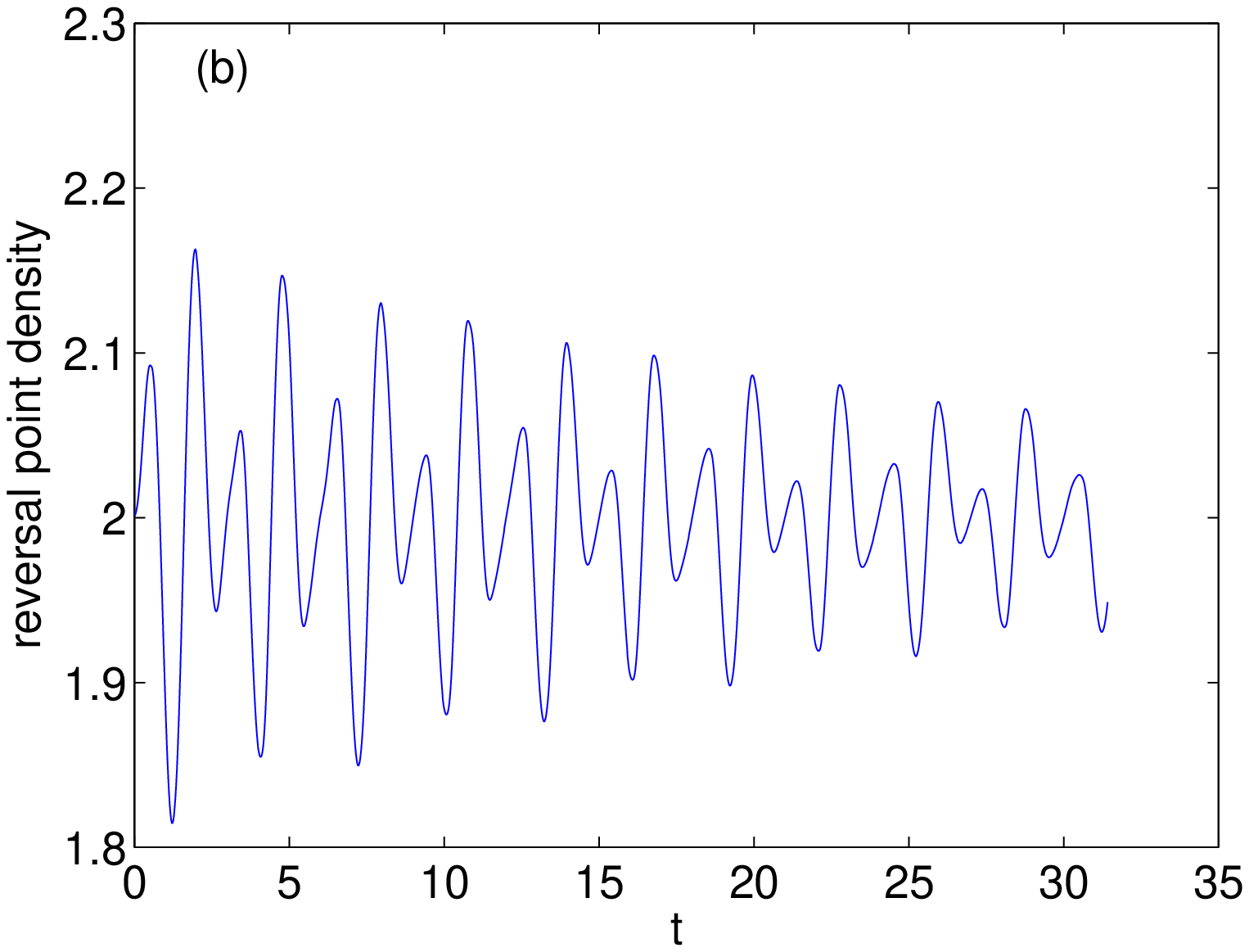} 
\includegraphics[width=8cm]{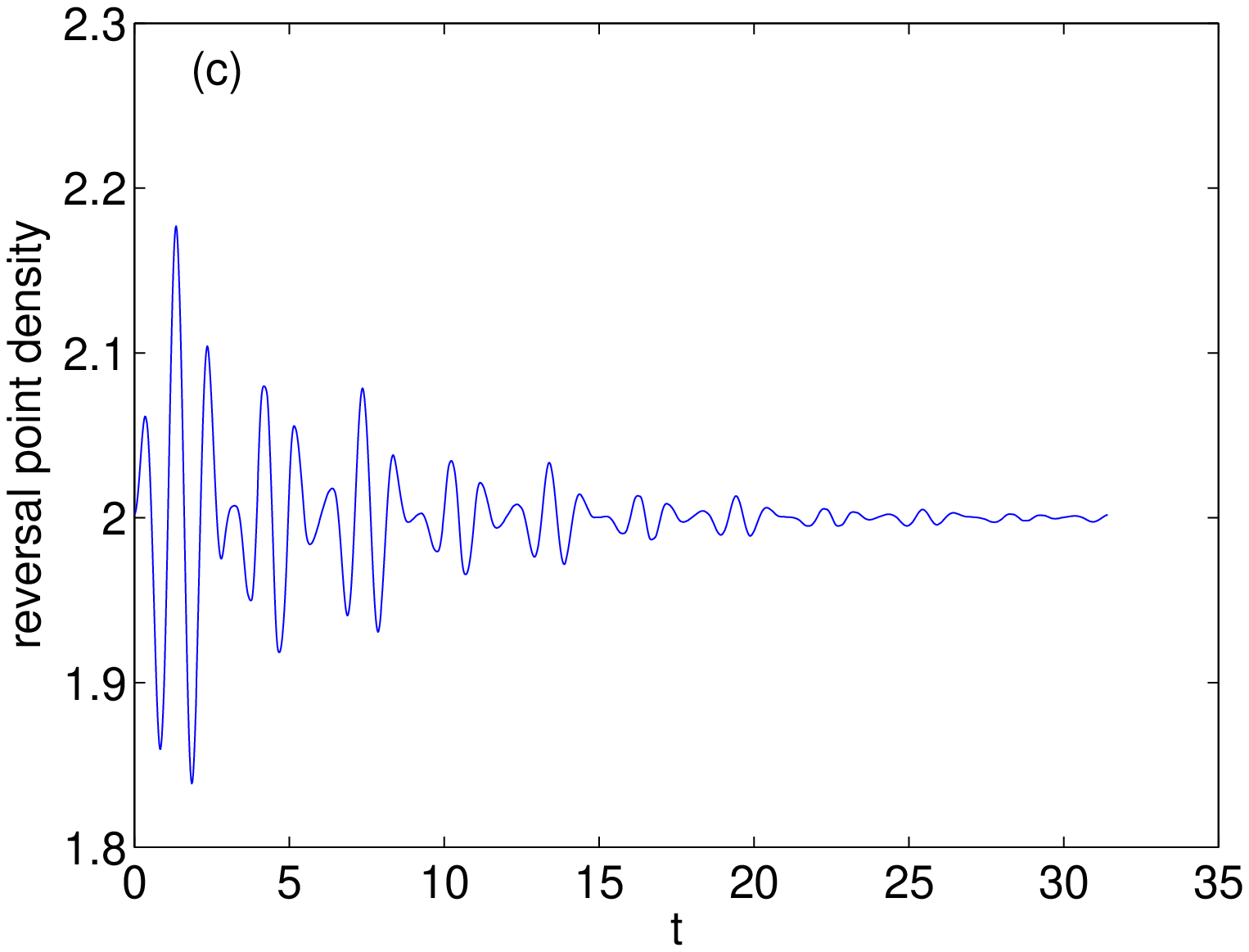} 
\end{center}
\caption{Plots of the RPD, $n_{\rm RPD}(x=0,t)$, vs.\ time for $\alpha=\pi/2$,  $\epsilon= 0.1$ and wavenumber: (a) $k=1$ (stable periodic pattern), (b) $k=4$ (transient to constant density), (c) $k=6$ (faster transient to constant density).  
\label{fig4}}
\end{figure}

The RPD exhibits the same behavior as the total density and its contour plot (not shown) is quite similar to it. In Figure \ref{fig4}, we depict the RPD at $x=0$ for the same values of the parameters. For unit wavenumber, periodic patterns are stable. The slight decrease of the maxima of the RPD observed in Fig.~\ref{fig4}(a) is due to unavoidable dissipation due to numerical errors. For $k=4$, the uniform stationary solution is stable according to the neutral stability curve of Fig.~\ref{fig1}. Fig.~\ref{fig4}(b) shows a pronounced evolution towards a constant. However, as we are not far from the bifurcation point (critical wavenumbers bounding the stability region for the uniform stationary solution are 3 and 5), the evolution of the RPD towards a constant value is slow. For $k=6$, neutral stability predicts periodic patterns. However at such large wavenumber, the dissipation due to numerical noise is so large that the pattern disappears and the uniform stationary solution becomes stable, as shown in Fig.~\ref{fig4}(c).
\bigskip 

\begin{figure}[ht]
\begin{center}
\includegraphics[width=8cm]{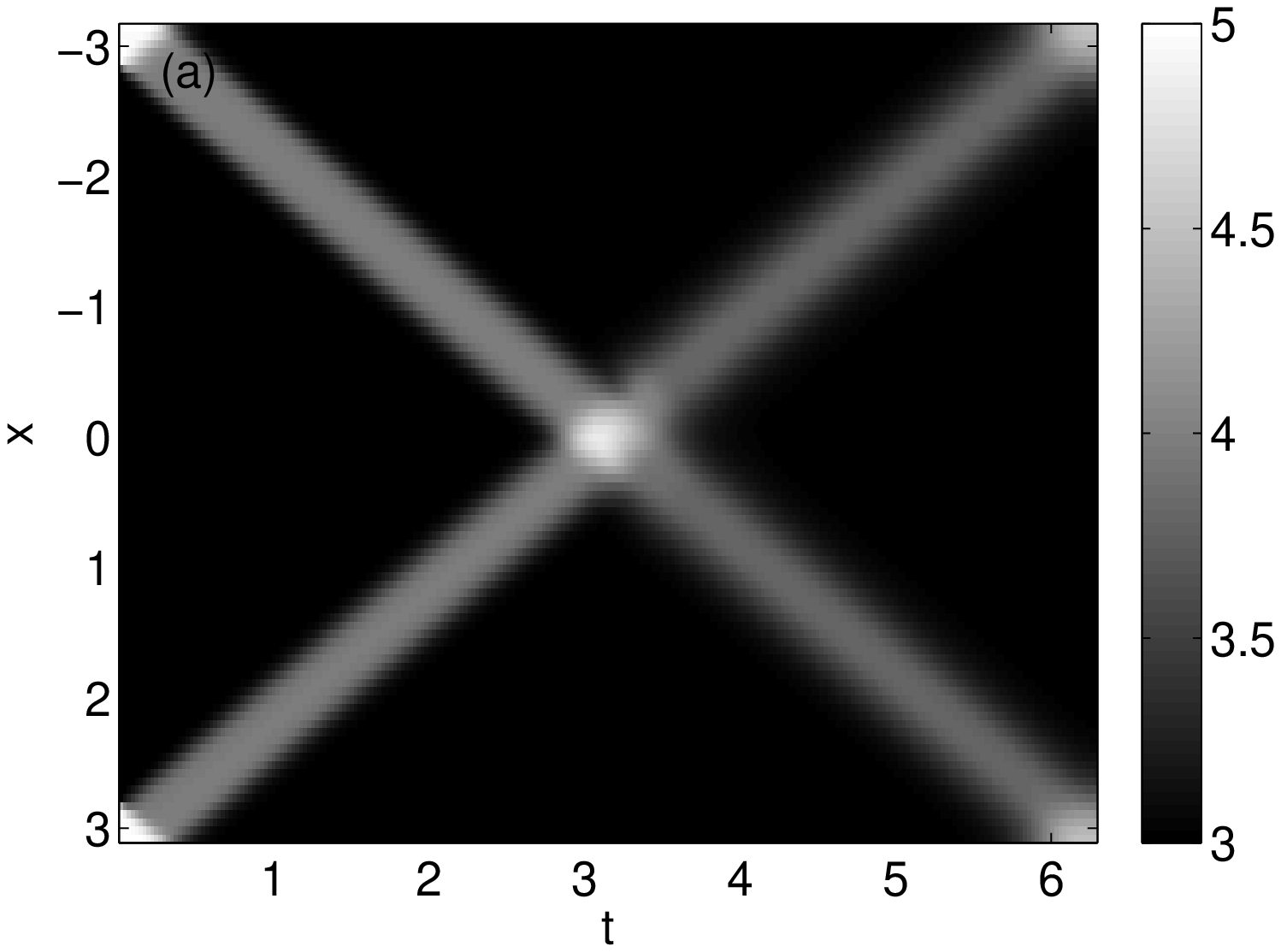} %\qquad
\includegraphics[width=8cm]{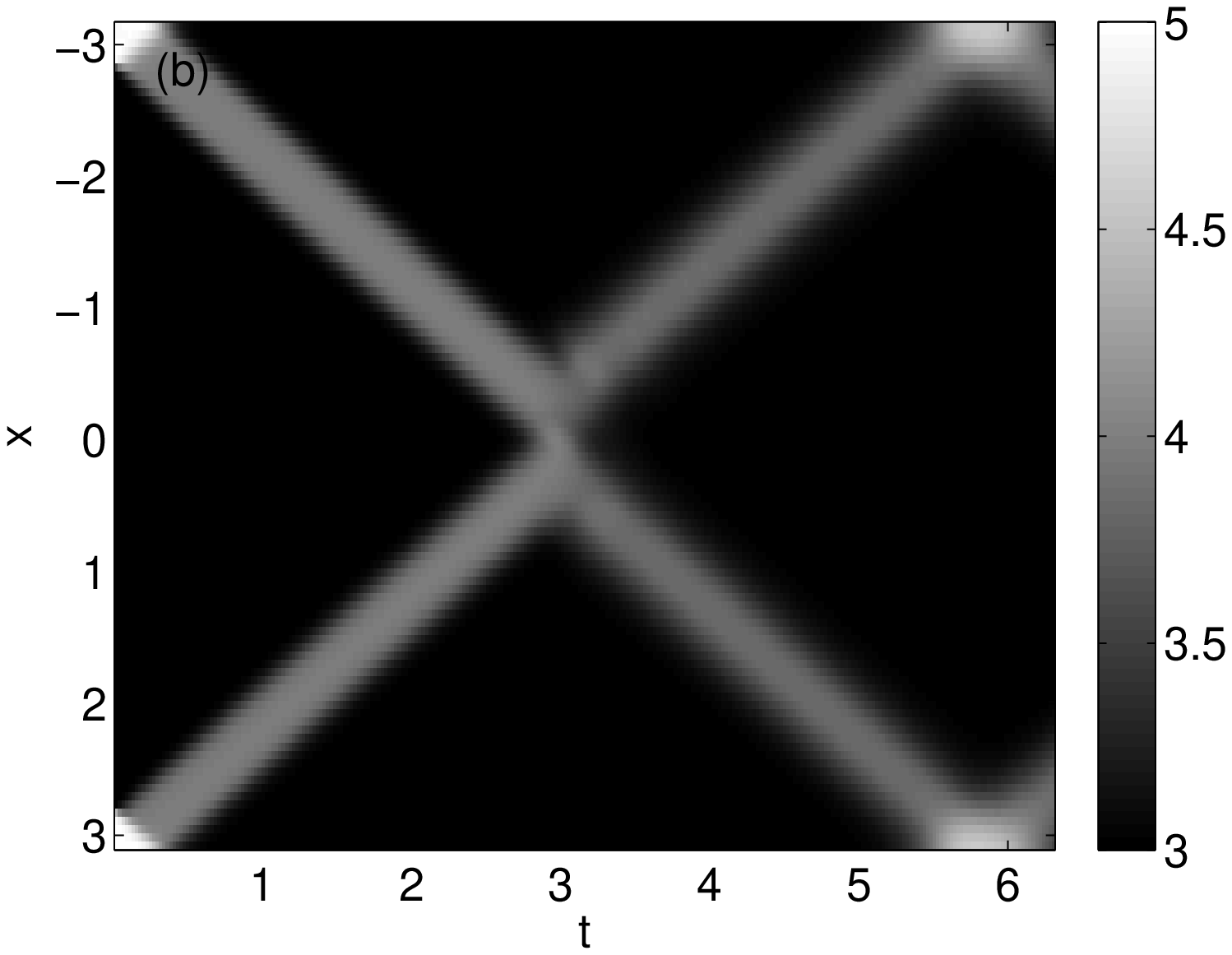} 
\includegraphics[width=8cm]{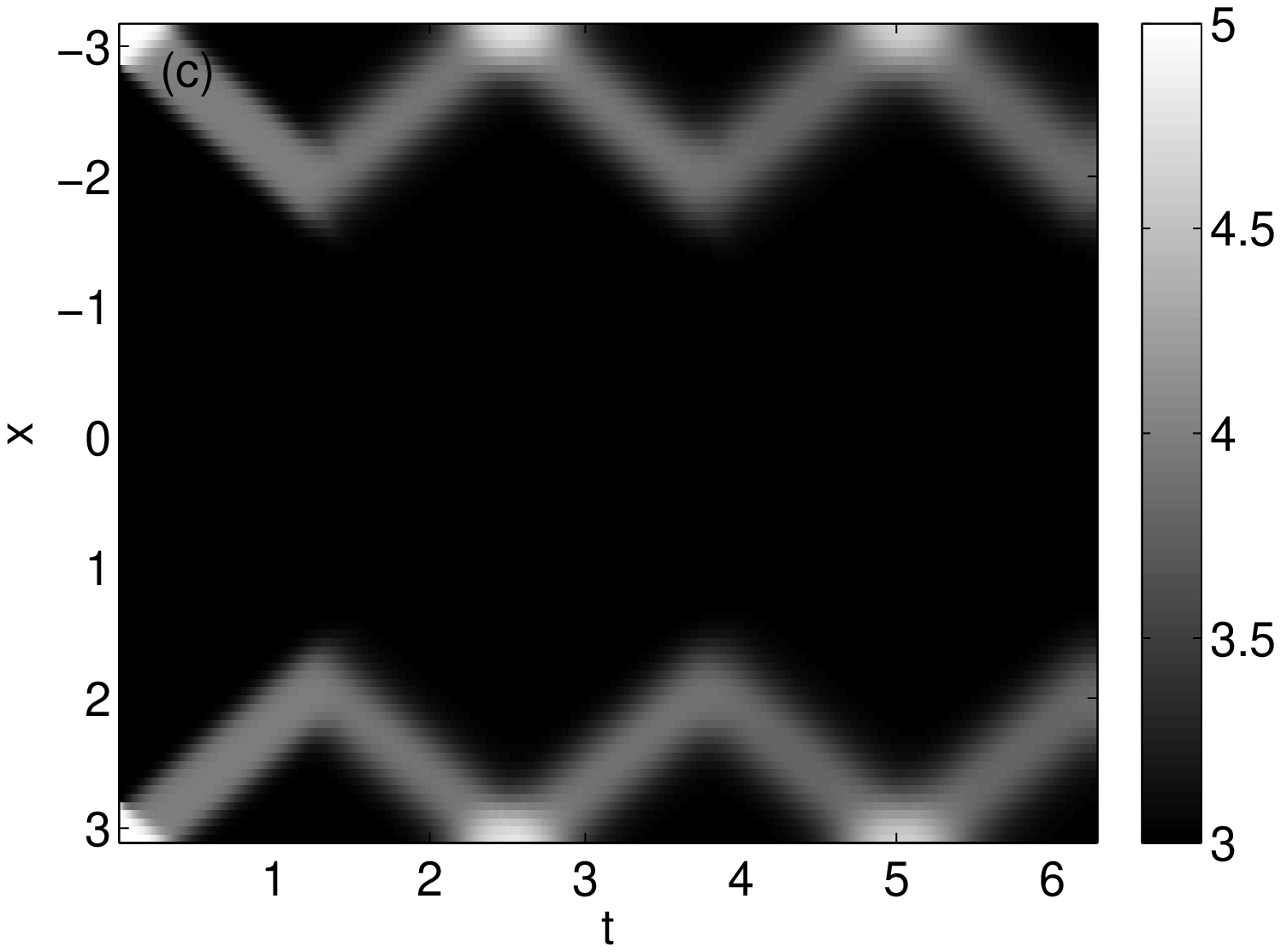} 
\includegraphics[width=8cm]{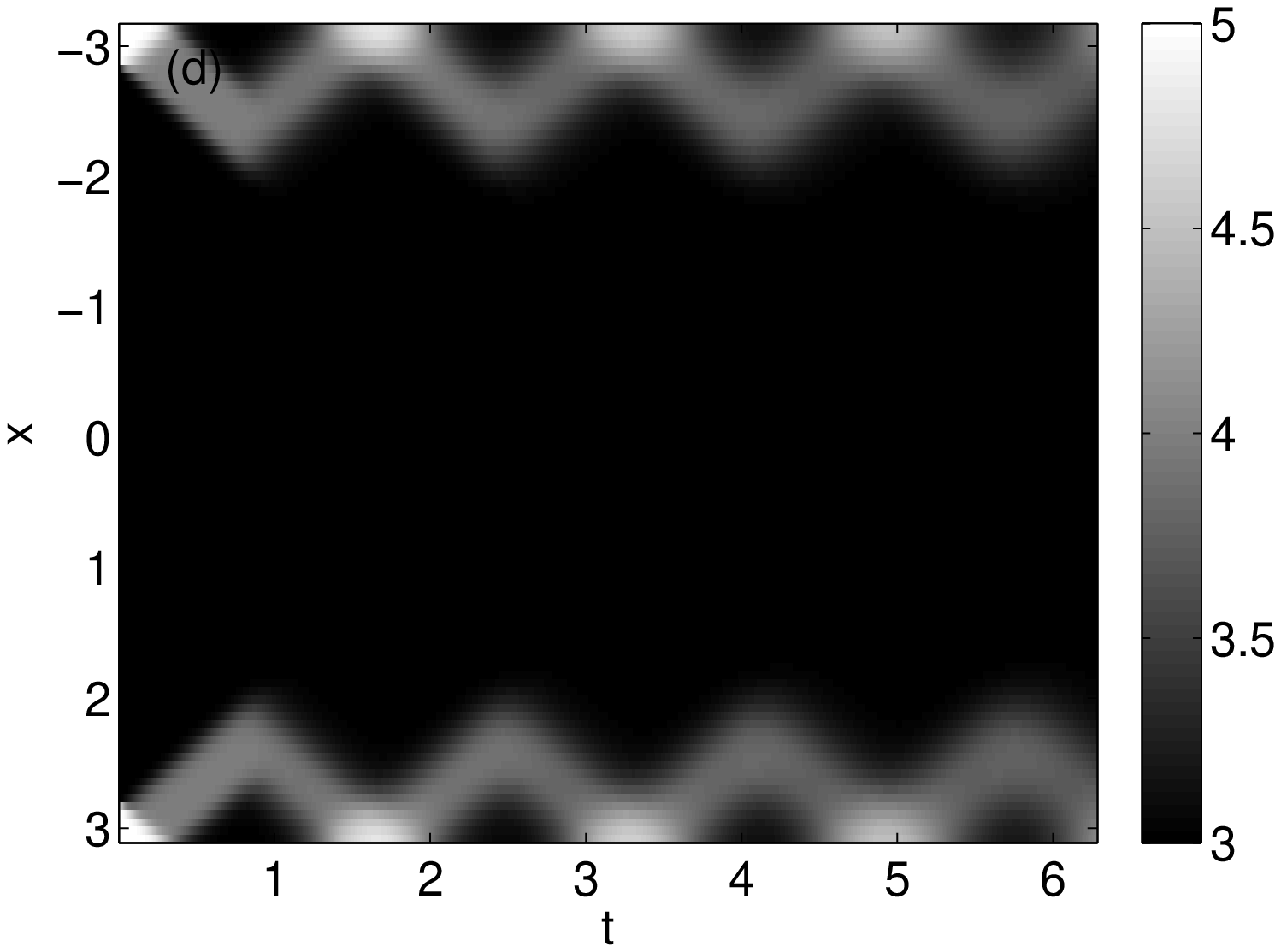} 
\end{center}
\caption{Contour plots of the RPD, $n_{\rm RPD}(x,t)$, for $\alpha=\pi/5$, initial condition  $n(x,\phi,0)=1.5 + [H(x-\pi+a)-H(x-\pi)] + [H(x+\pi)-H(x+\pi-a)]$, $a=\pi/10$, and: (a) $\epsilon= 0$, (b) $\epsilon= 0.1$, %1.2 
(c) $\epsilon= 3$, (d) $\epsilon= 12$. As $\epsilon$ (nonlinearity) increases, the reversals occur closer to $\pm\pi$.
\label{fig5}}
\end{figure}

Other patterns appear for different initial conditions. For example, for an initial condition with constant density for all $x$ except for two bumps near $x=\pm\pi$, $n(x,\phi,0)=1.5 + [H(x-\pi+a)-H(x-\pi)] + [H(x+\pi)-H(x+\pi-a)]$, $a=\pi/10$ [$H(x)=1$ for $x>0$, and $H(x)=0$ otherwise], we obtain the {\em standing wave patterns} shown in Fig.~\ref{fig5}. Reversals get confined to regions near $x=\pm\pi$ as the nonlinearity strength $\epsilon$ increases. For larger values of $\epsilon$, the weak signaling theory of Sections \ref{sec:3} and \ref{sec:4} does not apply. 

\begin{figure}[ht]
\begin{center}
\includegraphics[width=8cm]{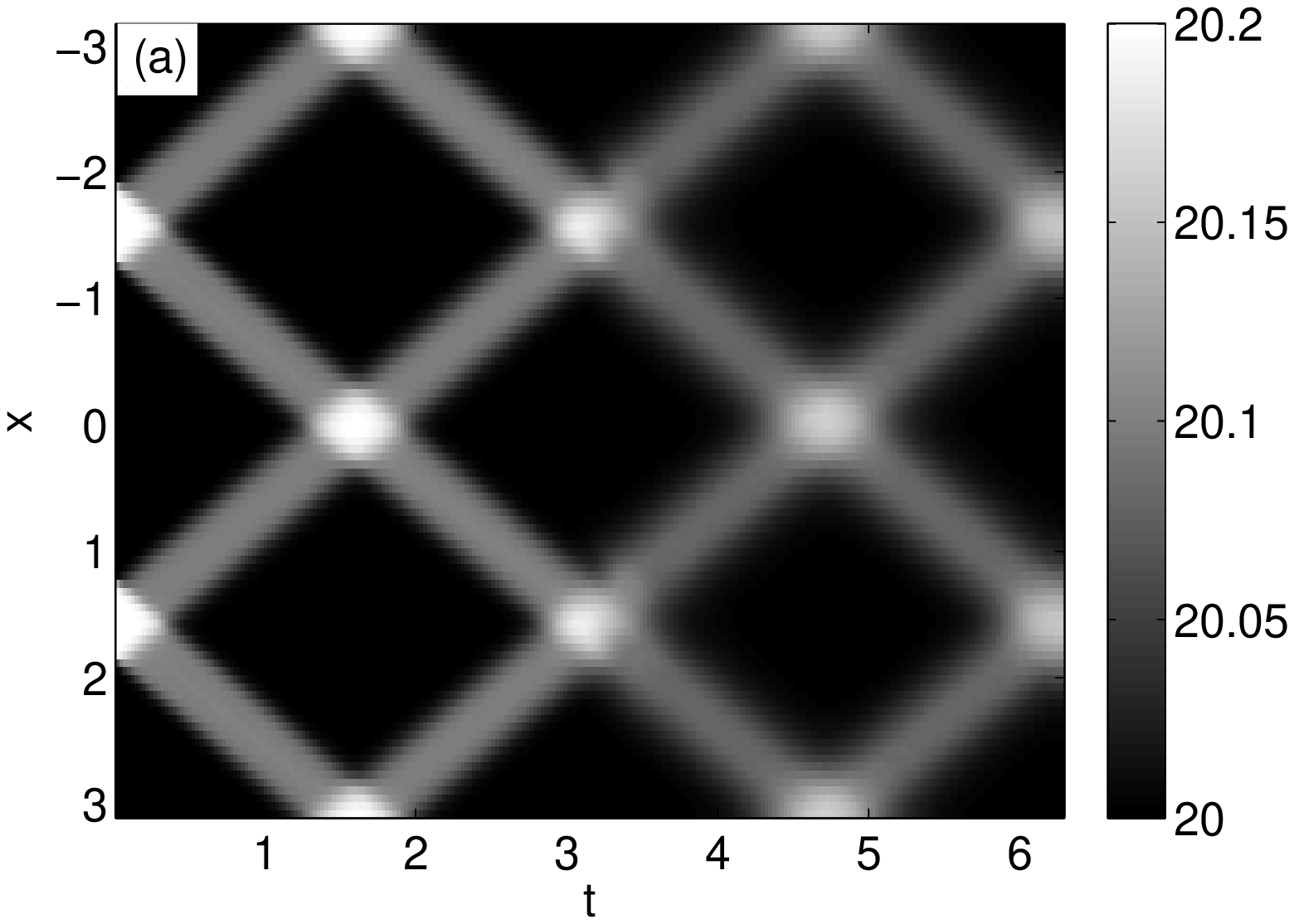} %\qquad
\includegraphics[width=8cm]{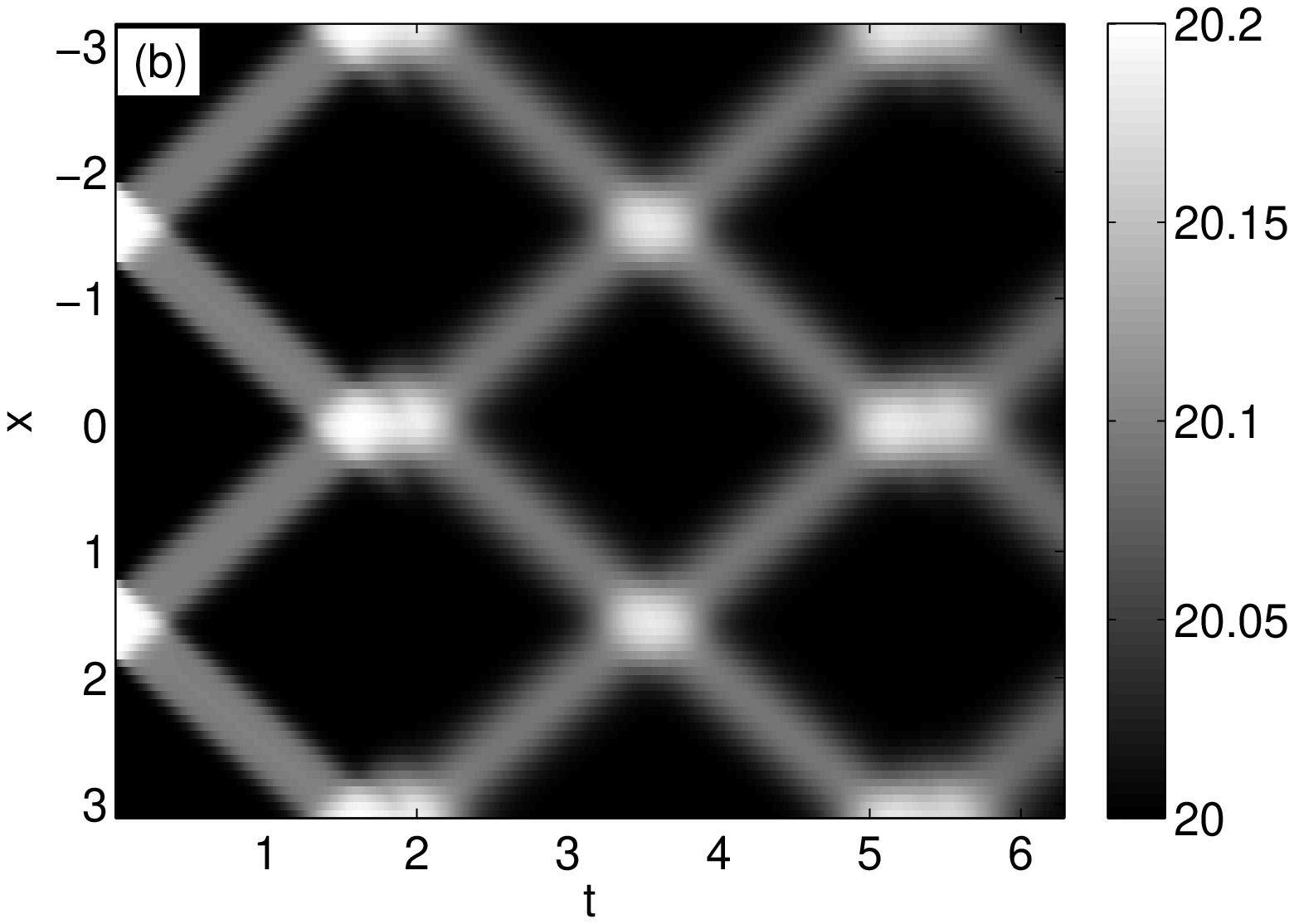} 
\includegraphics[width=8cm]{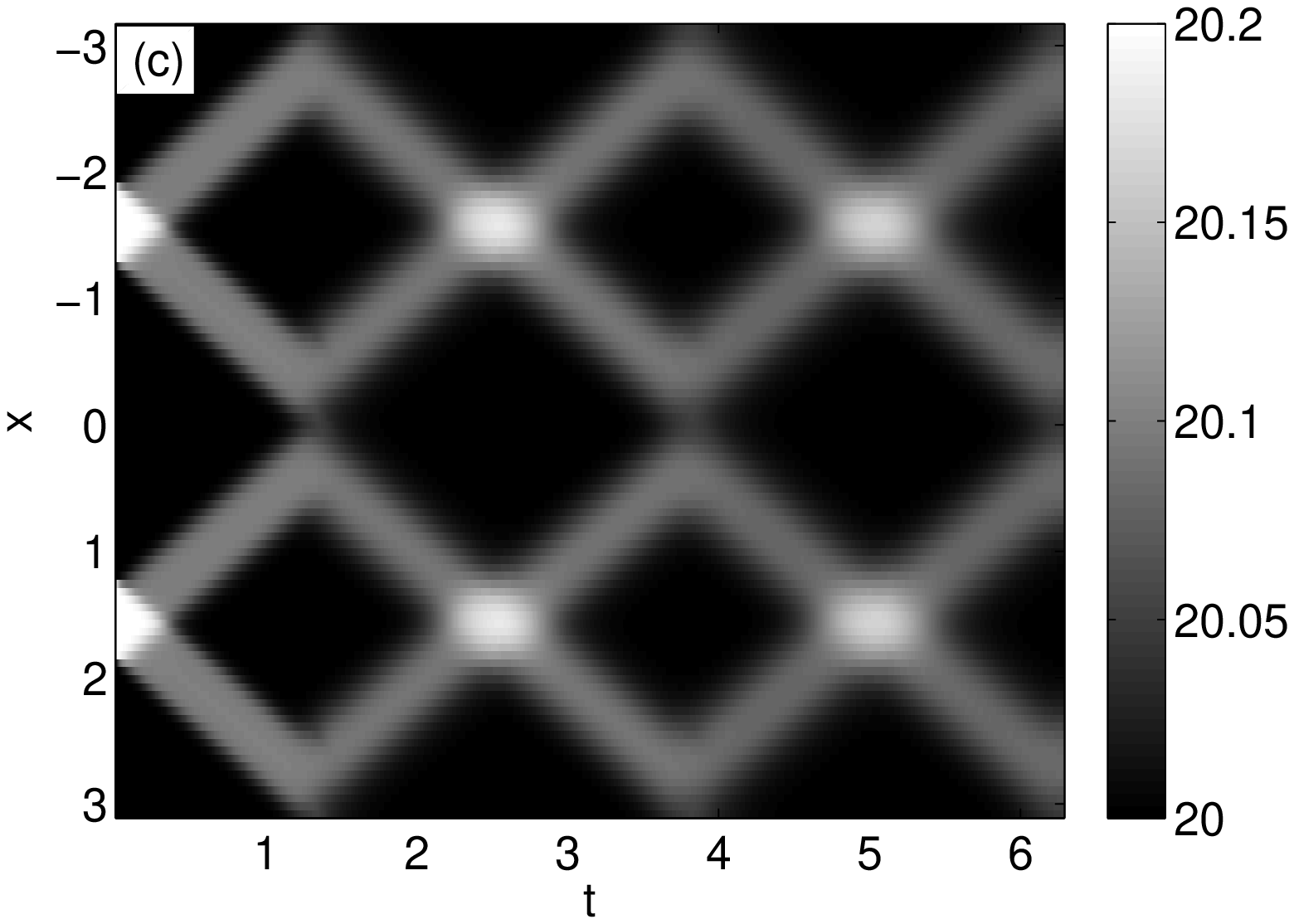} 
\includegraphics[width=8cm]{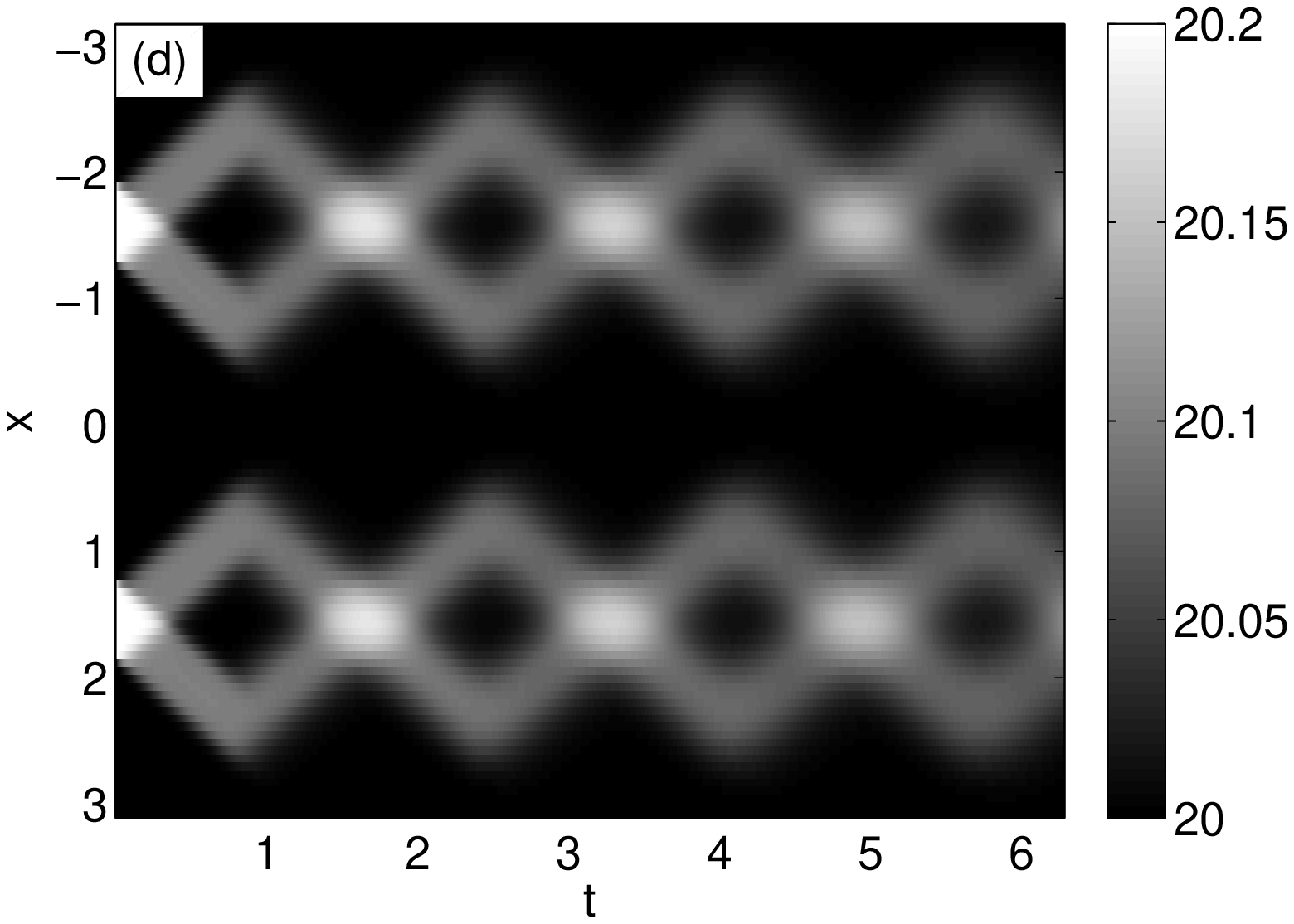} 
\end{center}
\caption{Contour plots of the RPD, $n_{\rm RPD}(x,t)$, for $\alpha=\pi/5$, initial condition  $n(x,\phi,0)=10 + 0.1\, [H(a-|x-\pi/2|)+H(a-|x+\pi/2|)]$, $a=\pi/10$, and: (a) $\epsilon= 0$, (b) $\epsilon= 1.2$, (c) $\epsilon= 3$, (d) $\epsilon= 12$. As $\epsilon$ (nonlinearity) increases, the reversals occur in smaller regions about $x=\pm\pi/2$.
\label{fig7}}
\end{figure}

If the two bumps are closer, then the standing waves generate local maxima as in Fig.~\ref{fig7}(a) and (b), for $\epsilon=0$ and 1.2, respectively. As $\epsilon$ (nonlinearity) increases, the reversals occur in smaller regions about $x=\pm\pi/2$, as shown in Figs.~\ref{fig7}(c) and (d). Clearly, for bumps that are close enough, the waves they issue reinforce the density at the points where they cross. In these points, fruiting bodies may form \cite{igo01}. It is interesting to observe that breaking the symmetry in the initial condition may weaken the resulting patterns. For example replacing the initial condition in Fig.~\ref{fig7} by $n(x,\phi,0)=1.5 + [H(x-\pi/2+a)-H(x-\pi/2)]+[H(x+\pi/2)-H(x+\pi/2-a)]$ (closer and narrower bumps, with centers at $\pi/2 -a/2$ and $-\pi/2+a/2$) results in patterns with two close maxima appearing near $t=n\pi$ ($n=1,2,\ldots$) that undergo noticeable dissipation.

\begin{figure}[ht]
\begin{center}
\includegraphics[width=8cm]{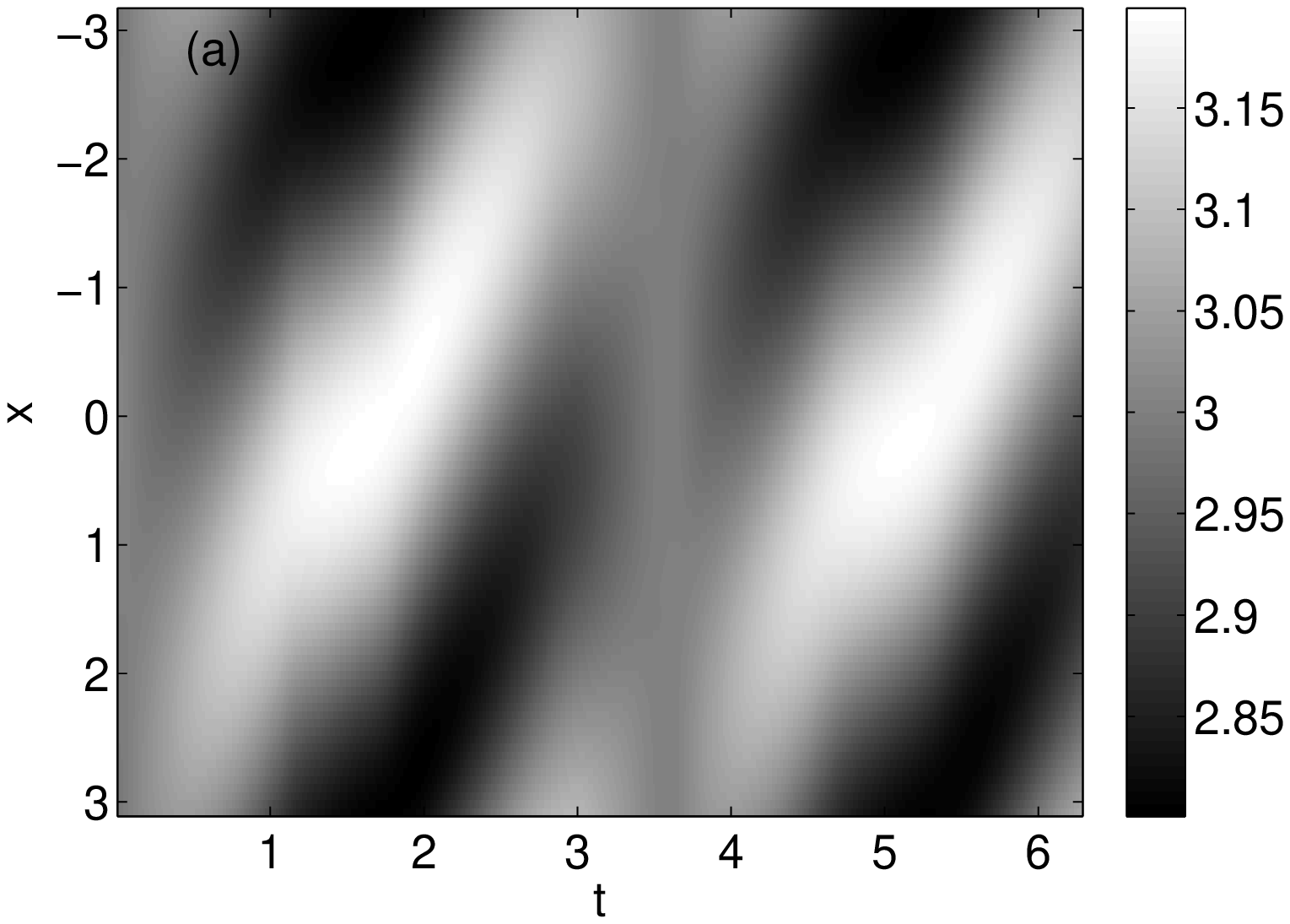} %\qquad
\includegraphics[width=8cm]{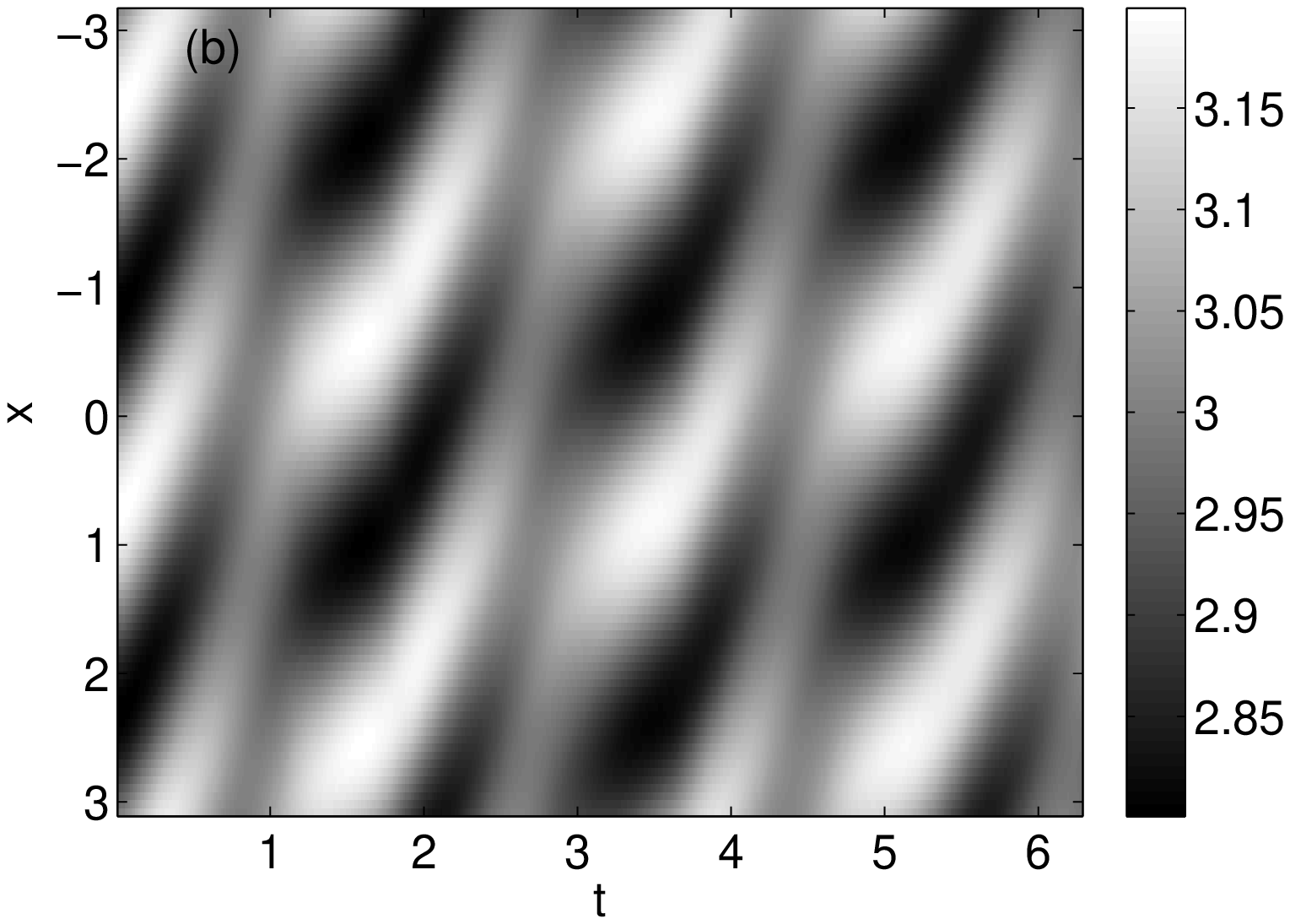} 
\includegraphics[width=8cm]{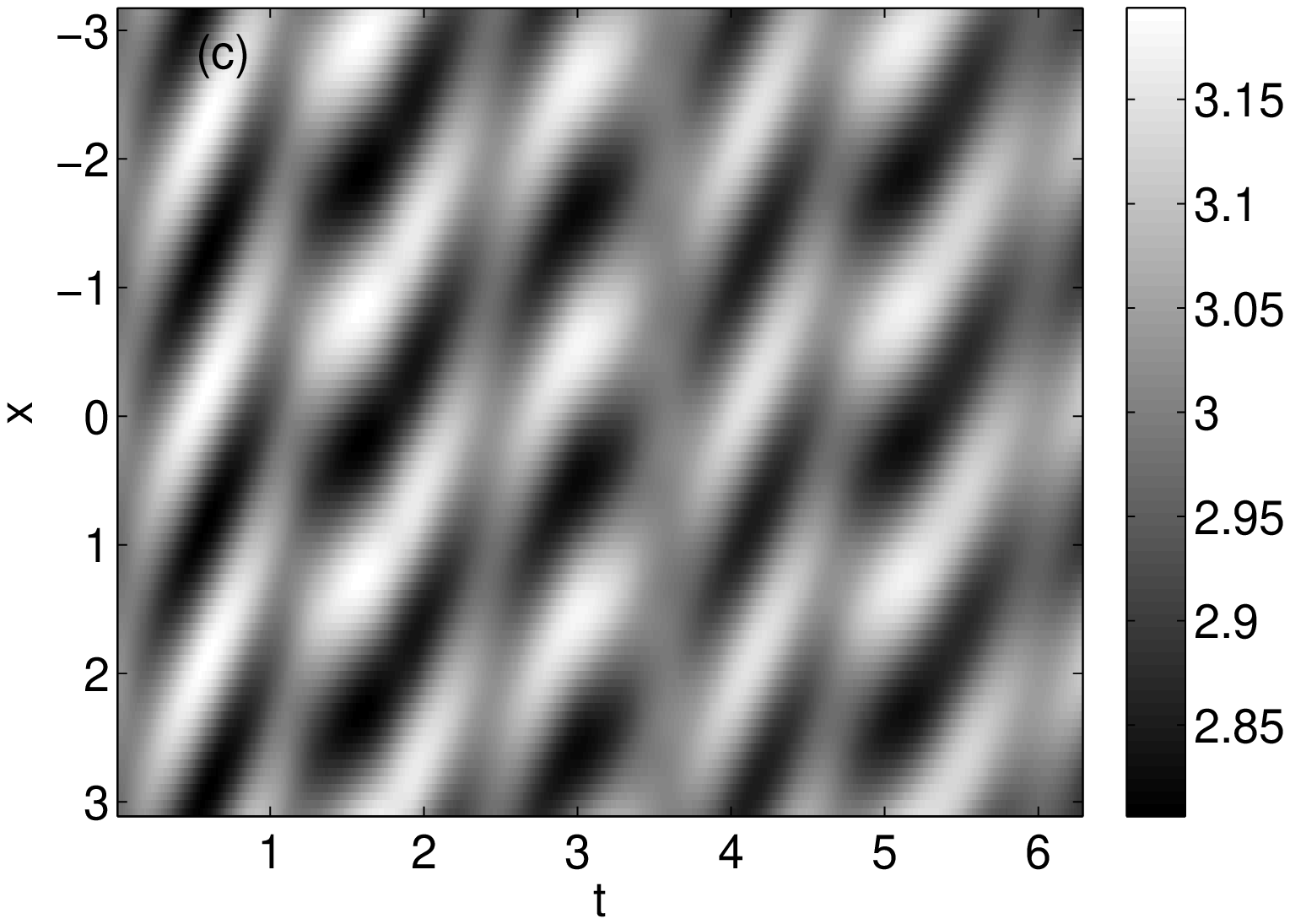} 
\end{center}
\caption{Contour plots of the RPD, $n_{\rm RPD}(x,t)$, for $\alpha=\pi/5$, initial condition  $n(x,\phi,0)=1.5 + 0.1\sin[k(x-\phi)]$, $\epsilon= 1.2$, and: (a) $k= 1$, (b) $k= 2$, (c) $k= 3$. As $k$ (wavenumber) increases, standing-wave time-periodic patterns become traveling-wave time-periodic patterns.
\label{fig6}}
\end{figure}

For an initial condition $n(x,\phi,0)=1.5 + 0.1\sin[k(x-\phi)]$, $\alpha=\pi/5$, the {\em standing wave patterns} shown in Fig.~\ref{fig6}(a) become the {\em traveling wave time periodic patterns} of Figs.~\ref{fig6}(b) and \ref{fig6}(c). 

\paragraph{Relation to patterns observed in experiments} The Igoshin  {\em et al} model we have solved numerically is 1D whereas observed patterns are 2D. This said, the complete patterns in panels (a) and (b) of Figures \ref{fig5} and \ref{fig7} are similar to those in Figure 3 of Welch and Kaiser's experiments \cite{wel01}, whereas loss of coherence (panels (c) and (d) of our Figures \ref{fig5} and \ref{fig7}) is observed in Figures 3(a) and 4 of the same work. Experiments also show more complex 2D patterns as in Figure 4 of \cite{bal15} that are reminiscent of our patterns in Fig.~\ref{fig6} although the 2D agent-based models the authors of \cite{bal15} introduce to explain the experiments are more sophisticated than the 1D model we study in this paper.

\section{Pattern decoherence and relation to the Kuramoto model}\label{sec:6}
The noiseless version of Igoshin et al's model (\ref{eq1})-\eqref{eq6} describes the density of myxobacteria with an internal clock in the limit as the number of bacteria ${\cal N}\to\infty$. The bacteria themselves satisfy the following equations
\begin{eqnarray}
&&\dot{x}_j=  v\mbox{ sign}\phi_j,\label{eq49}\\
&&\dot{\phi}_j= \omega+\epsilon\omega\,\Omega(N_{\rm sign(-\phi_j)}(x_j,t))[\chi_{[\alpha,\pi]}(\phi_j)+\chi_{[-\pi+\alpha,0]}(\phi_j)],\quad j=1,\ldots, {\cal N}, \label{eq50}\\
&&N_+(x,t)=\frac{\nu_s}{{\cal N}}\sum_{m=1}^{\cal N} \delta(x-x_m(t))\, H(\phi_m(t)),\quad N_-=\frac{\nu_s}{{\cal N}}\sum_{m=1}^{\cal N} \delta(x-x_m(t))\, H(-\phi_m).\label{eq51}%\\
%&&n_{\rm RPD}(x,t)=\frac{\nu_s}{{\cal N}}\sum_{m=1}^{\cal N} \delta(x-x_m(t))\, [\delta(-\phi_m(t)+0+) + \delta(-\pi-\phi_m(t)+0+) ].\label{eq52}% \nu_s=N.
\end{eqnarray}
Here $x_j(t)$ and $\phi_j(t)$ move on circles and can be considered to take values on the intervals $[-L,L]$ and $[-\pi,\pi]$, respectively, $\nu_s$ is a scaling parameter, and the delta functions are regularized in an appropriate way. In the limit as we take away this regularization and ${\cal N}\to\infty$, the densities \eqref{eq51} approach their continuum limits \eqref{eq5}. A typical bacterium $x_j(t)$ moves counterclockwise on a circle of length $2L$ if its internal phase $\phi_j(t)\in [-\pi,\pi]$ is positive, and clockwise otherwise. Its internal phase can be accelerated due to a mean-field interaction with opposite moving bacteria that collide with it. Appropriate initial conditions produce periodic (rhythmic) spatiotemporal patterns in the absence of interaction ($\epsilon=0$). Interaction tends to confine rhythmicity to parts of the circle $(-L,L)$, as shown in Figures \ref{fig5} and \ref{fig6}, or to destroy it. Loss of rhythmicity may appear as a {\em decoherence} phase transition at critical values of the refractory period $\alpha$ or the wavenumber of the initial condition, as shown in Figs.~\ref{fig3} and \ref{fig4}. The maxima of the time derivative of the reversal point density act as the order parameter for this phase transition: it is zero for the constant density solution and nonzero for the time-periodic patterns, see Fig.~\ref{fig4}(c). 

The behavior of this model can be compared with the well-known Kuramoto model of globally coupled phase oscillators \cite{kuramoto,rmp05}. In the Kuramoto model, the phases of free oscillators increase following their natural frequencies that are random (unsynchronized or incoherent state). Mean-field coupling between the oscillators succeed synchronizing them above a certain coupling strength and, typically, some oscillators are synchronized while others continue rotating about the unit circle (partial synchronization). In the Igoshin et al's model, a pattern induced by an appropriate initial condition persists in the absence of coupling. Turning on the coupling may confine the patterns to a part of the space interval $[-L,L]$ (partial decoherence) or destroy them completely (complete decoherence). Adding white noise sources, the Kuramoto model is described by a nonlinear Fokker-Planck equation whereas the Igoshin et al's model contain extra mechanisms of dissipation, as it is apparent from the nonlinear Fokker-Planck equation with an additional (collision) source term \eqref{eq34} obtained in the weak signaling limit. 

\section{Conclusions}\label{sec:7}
We have revisited the continuum model of rippling in myxobacteria proposed by Igoshin et al \cite{igo01,igo04}. In the absence of noise, the model consists of two coupled hyperbolic equations (describing the densities of left and right moving bacteria) coupled nonlinearly through a flux in an angular variable that represents the bacteria internal clock. This flux is a nonlinear function of the overall density of left or right moving bacteria. Depending on the values of the parameters, the model displays a variety of space and time periodic patterns that have been scarcely analyzed. 

In the limit of weak nonlinearity (weak signaling), we have found a Fokker-Planck type equation for the reversal-point density that contains a source term, absent in Igoshin et al's analysis \cite{igo04}. The reversal-point density can be used to reconstruct the densities of left and right moving bacteria. We analyze the linear stability of a constant-density solution and find that its neutral stability curves provide  selection rules giving the wavenumber of the patterns issuing from the constant-density solution. These selection rules issue directly from the source term in the Fokker-Planck equation. We have checked these results by direct numerical solution of the original hyperbolic equations of the model. For small nonlinearities, we have checked the wavenumber selection rule. Strengthening the nonlinearity tends to confine and destroy the patterns through a nonequilibrium phase transition. For large nonlinearity, we have found a variety of patterns including time-periodic standing and traveling waves that attest the richness of Igoshin et al's model. 

\acknowledgments
This work has been supported by the Ministerio de Econom\'\i a y Competitividad grants
FIS2011-28838-C02-01, MTM2014-56948-C2-2-P (LLB and AG) and MTM2014-56218-C2-2-P (AM). We thank Miguel Ruiz-Garcia (Universidad Carlos III) for fruitful discussions.

\end{document}